\documentclass[11pt]{article}
\usepackage{jheppub}

\usepackage{epsfig}
\usepackage{amssymb}
\usepackage{amsmath}
\usepackage{amsfonts}

\newcommand{\be}{\begin{equation}}
\newcommand{\ee}{\end{equation}}
\newcommand{\bea}{\begin{eqnarray}}
\newcommand{\eea}{\end{eqnarray}}
\newcommand{\bi}{\begin{itemize}}
\newcommand{\ei}{\end{itemize}}
\newcommand{\ben}{\begin{enumerate}}
\newcommand{\een}{\end{enumerate}}

\newcommand{\ep}{\epsilon}
\newcommand{\eps}{\epsilon}

\newcommand{\nn}{\nonumber}

\def\gsim{\mathrel{\rlap{\lower4pt\hbox{\hskip1pt$\sim$}}
    \raise1pt\hbox{$>$}}}         
\def\lsim{\mathrel{\rlap{\lower4pt\hbox{\hskip1pt$\sim$}}
    \raise1pt\hbox{$<$}}}         

\def \ep {\epsilon}

\begin{document}

\hfill{\today}

\newcommand{\picturepage}[2]{
\begin{minipage}{1cm}
      \includegraphics[width=30mm, height=30mm]{#1}
    \end{minipage} \quad &  
        \begin{minipage}[c]{15cm}    
            #2 
        \end{minipage}
   }

\title{Non-planar master integrals for the production of two off-shell 
vector bosons   in collisions of massless partons
}

\author[1]{Fabrizio Caola,}
\author[2]{Johannes M. Henn,}
\author[1]{Kirill Melnikov}
\author[4]{and Vladimir A. Smirnov}

\affiliation[1]{Department of Physics and Astronomy, Johns Hopkins University, Baltimore, USA
}
\affiliation[2]{Institute for Advanced Study, Princeton, NJ 08540, USA, USA}
\affiliation[4]{
Skobeltsyn Institute of Nuclear Physics of Moscow State University, 
119991 Moscow, Russia}

\emailAdd{caola@pha.jhu.edu}
\emailAdd{jmhenn@ias.edu}
\emailAdd{melnikov@pha.jhu.edu}
\emailAdd{smirnov@theory.sinp.msu.ru}

\abstract{
We present  the calculation  of all non-planar master integrals that are needed  
to describe production of two off-shell vector bosons in collisions of two massless 
partons  through NNLO in perturbative QCD. 
The integrals are computed analytically using 
differential equations  in external kinematic variables  and expressed 
in terms of Goncharov polylogarithms. 
These results  provide the last missing ingredient needed 
for the computation of  two-loop amplitudes that describe the production of two gauge 
bosons with different invariant masses  in hadron collisions.  
}

\maketitle

\section{Introduction} 

Perturbative QCD provides a viable framework to understand physics of hadron collisions.
Continuous progress with perturbative QCD computations was instrumental for the success
of the LHC physics program, crowned with 
the  celebrated discovery of the Higgs boson.  It is expected that 
a higher collision energy and the higher luminosity that will be reached when the LHC 
will resume its operations next year,  will enable  detailed studies of the multitude of various 
processes that involve elementary particles.  It is therefore important
to continue pushing  frontiers of perturbative QCD  in order to provide the 
best-possible  theoretical predictions for relevant physics observables.  A point in case is the production 
of two vector bosons, both on- and off-shell, in hadron collisions, $pp \to V_1^* V_2^*$.  
This process  is interesting for a variety of physics 
reasons that we recently summarized in~\cite{Henn:2014lfa}.  Considerations 
presented in~\cite{Henn:2014lfa}
strongly motivate the  extension 
of existing theoretical predictions  for this process 
\cite{dixon1,dixon2,bier,baglio,thr,Cascioli:2013gfa,shower,Campanario:2013wta}
to NNLO QCD. First and foremost, such an extension requires the scattering 
amplitude for a partonic processes $ ij \to  V_1^* V_2^*$  computed through two loops in perturbative QCD. 

In Ref.~\cite{Henn:2014lfa} we made a step towards the computation of this amplitude   by calculating 
all two-loop planar integrals  that contribute to these processes\footnote{
Results for planar master integrals for the case of vector bosons with equal masses were first
presented in~\cite{Gehrmann:2013cxs}.}. The goal of the current  paper 
is to complete the computation of the necessary ingredients for the 
two-loop amplitude calculation  by providing explicit results 
for all relevant {\it non-planar} integrals. To compute them, we use the method of differential 
equations as suggested in  Ref.~\cite{jhenn}. This  allows us to choose the master 
integrals in such a way that iterative solution  in the dimensional 
regularization parameter $\epsilon = (4-d)/2$  becomes straightforward.

The remainder of the paper is organized as follows. In the next Section,
 we introduce  our notation and explain the basic strategy.  In Section \ref{sec:diffeqs}
we discuss the differential equations and point out their general properties 
that are used later.  In Section \ref{sec:solution} we explain how we constructed 
the analytic solutions  of these differential equations  in terms of multiple polylogarithms 
in the physical region. We also discuss 
how boundary conditions are computed. 
In Section \ref{sec:masters}, we  list non-planar  master integrals and give their  boundary asymptotic behavior
in the physical region; we also present explicit results for divergences  of some integrals 
and describe checks of our results.
We conclude in Section \ref{sec:concl}.
Finally, in attached files, we give  matrices that are needed to construct the differential equations 
for our basis of master integrals and the analytic results  for all the non-planar two-loop three- and 
four-point integrals in terms of Goncharov polylogarithms. 

\section{Notation} 
\label{sec:notation}

We consider two-loop QCD corrections to  the
process $q(q_1) \bar q(q_2) \to V^*(q_3) V^*(q_4)$.  The four-momenta 
of external particles satisfy $q_1^2 =0, q_2^2 = 0$ and $q_3^2 = M_3^2$, $q_4^2 = M_4^2$. 
The Mandelstam invariants are\footnote{We use Mandelstam variables written with capital letters to refer 
to the physical process. Later, we will use Mandelstam variables for families of integrals; those we will write 
with lower case  letters.} 
\be
S = (q_1+q_2)^2 = (q_3 + q_4)^2,\;\;\; T = (q_1 - q_3)^2 = (q_2 - q_4)^2,\;\;\; U = (q_1 - q_4)^2 = (q_2 - q_3)^2;
\label{eq_man}
\ee
they satisfy  the standard constraint $S + T + U = M_3^2 + M_4^2$.   The physical values of these kinematic variables are 
$M_3^2 > 0, M_4^2 > 0$,   $S > (M_3 + M_4)^2$, $ T < 0$ and $U < 0$.  Further constraints on these variables can be derived 
by considering the center-of-mass frame of colliding partons and  expressing  the transverse 
momentum of each of the vector bosons $\vec q_\perp$ through $T$ and $U$ variables. We find
\be
\vec q_\perp^{~~2} = \frac{( T U - M_3^2 M_4^2)}{S}.
\ee
In addition, the square of the three-momentum of each of the vector bosons in the center-of-mass frame reads 
\be
\vec q\;^2 = \frac{S^2 - 2 S (M_3^2 + M_4^2) + (M_3^2 - M_4^2)^2}{4 S} \,.
\label{eqp}
\ee
The constraints on $T$ and $U$ for given $S, M_3^2, M_4^2$ follow from  obvious inequalities 
\be
 0 \le \vec q_\perp^{~~2} \le \vec q\;^2.
\ee

\begin{figure}[t]
  \centering
  \includegraphics[width=0.4\textwidth]{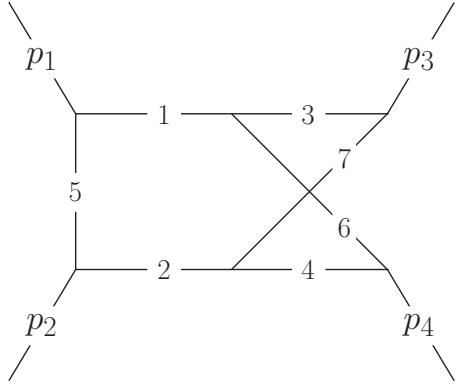}
  \caption{Double box non-planar  
graph. The numbering of the internal lines corresponds to the notation used in Eqs.~(\ref{eq_g}), (\ref{defpropagators}). The ingoing external momenta satisfy $\sum_i {p_i^\mu}=0$. 
  Different choices of on-shell conditions for them define the three 
non-planar integral families considered in the main text.}
  \label{figuredoublebox}
\end{figure}

All non-planar two-loop diagrams that are required for the production 
of two off-shell vector bosons
can be   described by a 
single meta-graph shown in Figure \ref{figuredoublebox}.  Three mappings, that define three 
distinct families of integrals, need to be considered:
\begin{enumerate} 
\item family $N_{12}$:  $ p_1 = -q_4,\; p_2 = -q_3,\;  p_3 = q_2,\;  p_4 = q_1$;
\item family $N_{13}$:   $ p_1 = -q_4,\;  p_2 = q_2,\;  p_3 = -q_3,\;  p_4 = q_1$;
\item family $N_{34}$:   $ p_1 =  q_1,\ p_2 = q_2,\; p_3 = -q_3,\; p_4 = -q_4$.
\end{enumerate} 

For each of these families, we define a set of integrals that is closed under the application of 
integration-by-parts  identities. Specifically, 
\be 
G_{a_1,\ldots,a_9}=
\int \frac{{\rm d}^D k_1}{ i \pi^{D/2} } 
\frac{{\rm d}^D k_2}{ i \pi^{D/2} }
\frac{1}{[1]^{a_1} [2]^{a_2} [3]^{a_3} [4]^{a_4} [5]^{a_5} [6]^{a_6} [7]^{a_7} [8]^{a_8} [9]^{a_9}},
\label{eq_g}
\ee
and 
\be
\begin{split} \label{defpropagators}
& [1] = -k_1^2,\;\;\; [2] = -(k_1+p_1+p_2)^2,\;\;\; [3] = -k_2^2,\; \;\; \\
& [4] = -(k_1 - k_2+p_1+p_2 + p_3)^2,\;
[5] = -(k_1+p_1)^2, \;\;\; [6] = -(k_1 -k_2)^2, \\
& [7] = -(k_2-p_3)^2,\;\;\;\; [8] = -(k_2 + p_1)^2,\;\;\;\;\; [9] = -(k_1-p_3)^2.
\end{split} 
\ee
The exponents can take any integer values,  with the restriction that $a_8 \le 0$ and $a_9\le 0$.
For each of the three families, 
integration-by-parts identities can be used to express all the integrals of that type through a minimal 
set of (master) integrals. 
Our choice of master integrals can be found  in ancillary  files. Many of these master integrals 
are, in fact, planar  and were computed by some of us in Ref.~\cite{Henn:2014lfa}.
Genuine non-planar master integrals for each of the three families  are shown and 
discussed in Section~\ref{sec:masters}.

All master integrals  satisfy differential equations in the external 
kinematic variables. These differential equations can be simplified by choosing 
suitable parametrizations of kinematic invariants.  For the families $N_{12}$ and $N_{13}$ 
we choose the parametrization to be 
\be
S = M^2 (1+x) ( 1 + x y),\;\;\; T = -M^2 xz,\;\;\; M_3^2 = M^2,\;\;\;M_4^2 = M^2 x^2 y,
\label{eq_xyz}
\ee
where $M^2$ is the overall scale parameter.  
We note that the above parametrization is the same as in  the planar case \cite{Henn:2014lfa}
and that in terms of the variables $x,y,z$, the physical region corresponds to 
\be
x > 0,\;\;\; y > 0,\;\;\;\; y < z < 1. 
\ee

For the family $N_{34}$, the above parametrization can also be used but it is not optimal 
since 
it leads to the appearance of multiple letters in an alphabet that are quadratic in $x,y$ and $z$ which 
is problematic for the construction of an analytic solution that is based on Goncharov polylogarithms. 
Instead, we find it  useful to choose the following parametrization
\be
\begin{split} 
& S = M^2 (1+x)^2, 
\;\;\; T = -M^2 x \left ( ( 1+ y) (1 + xy) - 2z y (1+x) \right), \\
& M_3^2 = M^2 x^2(1-y^2),\;\; M_4^2 = M^2 (1-x^2 y^2).
\label{eq2p9}
\end{split}
\ee
While the above parametrization also does not lead to a linear alphabet, it allows 
us to construct solutions in terms of Goncharov polylogarithms, as we explain below.\footnote{We note 
that it is possible to obtain a linear alphabet for the N34 family by changing variables 
$x \to x/y$ in Eq.(\ref{eq2p9}).}

\section{Differential equations} 
\label{sec:diffeqs}

In this Section we describe a procedure~\cite{jhenn} that allows us to compute the master 
integrals and comment on some aspects that arise when this procedure is applied to the 
calculation of non-planar integrals.  We begin by deriving 
  systems of differential equations for each of the above families. 
This is a relatively standard procedure, see e.g. Refs.~\cite{kotikov,remiddi}, and we do not discuss 
it further. When deriving differential equations we performed a reduction to master integrals
using the {\tt c++} version of program {\tt FIRE} \cite{Smirnov:2008iw,Smirnov:2013dia}.
We choose all master integrals to be dimensionless, such that they depend only
on the three variables $x,y,z$ and find 
\be
\partial_\xi \vec f = \epsilon {A}_\xi \vec{f},
\label{eq_f}
\ee
where  $\xi = x,y,z$ and $\vec{f}$ is a vector of master integrals.
The master integrals for all the three families can be found in ancillary files;
some examples of master integrals are discussed in Section~\ref{sec:masters}.

The matrices ${A}_\xi$ contain simple rational functions.
They satisfy the integrability conditions
\be
\left (  \partial_\xi \partial _\eta - \partial_\eta \partial_\xi \right ) \vec f =0 \;\;\;\;
\Rightarrow\;\;\;\;\; 
\partial_{\xi} A_{\eta} - \partial_{\eta} A_{\xi} = 0\,,\qquad [A_{\eta} , A_{\xi} ] = 0\,,
\ee
for $\xi,\eta \in \{x,y,z\}$.
The structure of the equations can be further clarified by writing them in the
combined form
\be \label{DEdifferentialform}
d\, \vec{f}(x,y,z;\eps) = \epsilon\,  d \, \tilde{A}(x,y,z) \, \vec{f}(x,y,z;\eps) \,,
\ee
where the differential $d$ acts on $x,y$ and $z$. Matrices 
${\tilde A}$ for each of the three families can be found in the ancillary files as well. 
For our choice of master integrals,
the matrix $\tilde{A}$ can be written in the following way
\begin{align}\label{Atilde}
\tilde{A}  = \sum_{i=1}^{N_{\rm max}}  \tilde{A}_{\alpha_i} \, \log(\alpha_{i} ) \,,
\end{align}
where the $ \tilde{A}_{\alpha_i}$ are {\it constant} matrices, and 
the arguments of the logarithms $\alpha_{i}$, called {\it letters}, are simple
functions of $x,y,z$. The length of the alphabet $N_{\rm max}$ depends on the integral family. 
For families  $N_{12}$ and $N_{13}$, we find the alphabet to be\footnote{First fourteen letters 
in Eq.(\ref{eq_alphabet1}) give the alphabet for $N_{12}$.}
\be
\begin{split}
\alpha_{N_{12} \& N_{13}}  = & \{ x, 1 + x, 1 - y, y, 1 + x y, 1 + 
 x (1 + y - z), 1 - z, z - y, 1 + y - z,     \\
& 1 + y + x y - z, z, 
x y + z, 1 + x + x y - x z, 1 + x z,  \\
& 1 + y + 2 x y - z + x^2 y z, z - y (1 - z - x z) \}.
\end{split}
\label{eq_alphabet1}
\ee
For the family  $N_{34}$, the alphabet reads 
\be
\begin{split}
\alpha_{N_{34}} = & \{ 
x, 1 + x, 1 - y, y, 1 + y, 1 - x y, 1 + x y, 1 -
 y (1 - 2 z), 1 + y - 2 y z, \\
& 1 - x y^2 - 
 y (1 - x - 2 z + 2 x z), 1 - x y (1 - 2 z), 1 + x (y - 2 y z), 
\\
& 
1 + 
 x y^2 - (1 + x) y (1 - 2 z), 1 - z, z, 1 + y - 2 y z, (1 + y)(1 + xy) -2z y(1+x),
 \\
& 1-y + 2yz,
 1 - 
 x y^2 + (1 - x) y (1 - 2 z) \}.
\end{split} 
\label{eq_alphabet2}
\ee

There are two things to be said about these alphabets for non-planar families. First, in contrast 
to planar master integrals, 
these alphabets contain quadratic polynomials. However,  thanks to the chosen parametrization, 
for each integral family there is just one variable ($x$ for $N_{12},~N_{13}$ and $y$ for $N_{34}$) 
with respect to which a particular alphabet is quadratic.   Constructing 
explicit solutions for non-planar integrals requires integrating these alphabets 
over $x,y$ and $z$. 
This is not easy to do if quadratic polynomials need to be integrated. 
Nevertheless, it turns out that these alphabets can be integrated 
without much trouble provided that we  postpone integration 
over quadratic variables until the very end of the calculation. 
Using this approach,  integration can easily be performed in terms of Goncharov 
polylogarithms.  We discuss this in more detail in the next Section.

Second, we note that  in the physical region, alphabets in Eqs.(\ref{eq_alphabet1},\ref{eq_alphabet2}) 
are sign-definite. 
This feature implies that all iterated integrals needed 
for calculating $\vec{f}$ can be written in a manifestly real form, so that  imaginary parts appear 
only through explicit factors of $i$. The latter come from the boundary 
conditions when they are computed in  the physical region.  This feature is similar to the 
case of planar master integral recently discussed in \cite{Henn:2014lfa}.

\section{Solution in terms of multiple polylogarithms} 
\label{sec:solution}

In this Section we review the procedure that allows 
us to solve the differential equations for  the master integrals, following the  
discussion in our previous paper \cite{Henn:2014lfa}.
The vector of master integrals $\vec f$ can be expanded in powers of $\epsilon$, 
\be
\vec f = \sum \limits_{i=0}^{4} \vec f^{(i)} \epsilon^i +  {\mathcal{O}}(\eps^5).
\ee
To construct a solution of the differential equation, we need to iteratively solve Eq.~(\ref{eq_f}) order-by-order 
in the 
dimensional-regularization parameter $\epsilon$.  Suppose the solution 
is constructed up to $i = n-1$. The set of differential 
equations for $\vec f^{(n)}$ is then 
\be
\partial_x \vec f^{(n)} = A_x \vec{f}^{(n-1)}, \;\;\; \partial_y  \vec{f}^{(n)} =  A_y \vec{f}^{(n-1)},
\;\;\; \partial_z  \vec{f}^{(n)} =  A_z \vec{f}^{(n-1)}.
\label{eq_de}
\ee
To find $\vec f^{(n)}$, we integrate the first equation over $x$; substitute the solution back to the 
differential equation for $y$, integrate again, substitute the solution back into the differential 
equation for $z$ and integrate again. This procedure determines $\vec f_n$  up to a constant of integration
 that is then fixed  from boundary conditions. 

To solve the  differential equations in Eq.(\ref{eq_de}), we should be able to integrate 
inverse elements of the alphabets that we displayed  in the previous Section. 
Since, as we pointed out already, elements of the alphabet can be both linear and quadratic 
in certain variables, such integration appears to be more complicated than the case of a linear 
alphabet that always permits to write a solution in terms of 
Goncharov polylogarithms, 
\be
G(a_n,a_{n-1},....a_1,t) = \int \limits_{0}^{t} \frac{{\rm d} t_n }{ t_n -a_n }
G(a_{n-1},....a_1,t_n). 
\label{eq_gn}
\ee
This concern is, however, unfounded and solution in terms of Goncharov polylogarithms 
can be constructed for the alphabets of families $N_{12}$, $N_{13}$ and $N_{34}$   as well. To show this, we consider, 
for definiteness, the alphabet of the family  $N_{34}$ which is quadratic 
in $y$ but is linear in $x$ and $z$.   
Suppose we choose to integrate over $z$ first, over $x$  second, and, finally, over $y$.
One of the quadratic  letters of the alphabet $\alpha_{34}$ reads
$1 + x y^2 - (1 + x) y (1 - 2 z)$. Upon integrating it over $x$, we obtain 
Goncharov polylogarithms of the form $\displaystyle G[(1-y+2 yz)/(y (1-y-2z)),...,x]$.
When this solution is used to derive the differential equation for the function 
of the other two variables $z,y$, 
all functions that depend on $x$ should cancel out. This implies 
that, by the time we get to the $y$-integration, 
 all the letters of the alphabet that are quadratic in $y$ and, at the same time, depend on 
$z$ and $x$ should disappears. Since there are no letters in the alphabet that are quadratic in $y$ and 
are independent of $x$ and $z$, we conclude that 
only letters of the alphabet that are linear in $y$ appear  at  the final 
stage of the integration. A similar consideration  shows that, in case of the family $N_{13}$,
we can avoid the need to deal with quadratic letters of the alphabet provided that 
we  first integrate over $y$, then over $z$ and, finally, over $x$.

A full construction of analytic solutions requires boundary conditions and 
we compute  them  directly 
in the physical region. Because of the different parametrizations used for different 
integral families,  boundary 
conditions are obtained  from different limits of $x,y$ and $z$ variables. 
For families $N_{12},N_{13}$, we consider 
the limits $x \to 0, y \to z, z \to 1$, and for the family $N_{34}$, the limit $x \to 0, y \to 0$ and 
$z \to 1$. Note that the physical meaning of these limits corresponds to the kinematic 
situation where the colliding partons have just enough energy to produce two vector bosons with 
very different masses $M_i^2 \ll M_j^2 \sim S$.
 In this  limit,  the absolute value of the 
three-momenta of vector bosons vanishes and the scattering occurs in the forward direction. 
This is the same kinematics that 
we used in our previous paper on planar master integrals~\cite{Henn:2014lfa} and, since 
many planar integrals appear in the current computation  as non-homogeneous 
terms in the differential equations, the boundary conditions computed in \cite{Henn:2014lfa} 
can be recycled for a large number of required integrals.  

The boundary conditions are {\em a priori} unknown for genuine non-planar integrals and we compute them 
in two different ways.  One possibility is to study consistency conditions of differential 
equations in three variables; this procedure is discussed in Ref.~\cite{Henn:2014lfa} and 
it is often sufficient to fix the  required boundary behavior of non-planar  integrals. 
Another possibility is to compute the relevant  limits directly, expanding Feynman integrals 
in small kinematic variables.  To accomplish this, we used  the strategy of expansion by 
regions \cite{Beneke:1997zp,Smirnov:2002pj} (for a recent review see Chapter~9 of Ref.~\cite{Smirnov:2012gma})
and its implementation in the public computer code {\tt asy.m} \cite{Pak:2010pt,Jantzen:2012mw}
which is now included into {\tt FIESTA} \cite{Smirnov:2013eza}.

\section{Master integrals} 
\label{sec:masters}

For each family of integrals, the Mandelstam variables are given by $s = (p_1+p_2)^2 = (p_3+p_4)^2$, 
$t = (p_1+p_3)^2 = (p_2+p_4)^2$, $u = (p_2+p_3)^2 = (p_1+p_3)^2$. Their relation 
to the physical Mandelstam 
variables $S,T,U$ and the ensuing parametrization in terms of variables $x,y,z$ can be read off using 
the $q \to p$ mapping just before Eq.~(\ref{eq_g}) and Eqs.~(\ref{eq_man}), (\ref{eq_xyz}).

We  choose the master integrals following  the strategy suggested  in Ref.~\cite{jhenn}.
The idea is to find master integrals whose Laurent expansion in $\ep$ leads to expressions 
of a uniform weight.  As guiding principles for  finding such integrals,
we analyzed generalized unitarity cuts, as well as explicit (Feynman) parameter representations 
of the integrals.  
Technically this is very similar to the analysis of certain three-loop massless integrals studied in 
Refs. \cite{Henn:2013tua,Henn:2013nsa}. In fact, some of the two-loop integrals 
with two off-shell legs are contained in those three-loop integrals as subintegrals. 
For more detailed explanations and examples, see Section 2 of Ref.~\cite{Henn:2013tua}. In 
Ref.~\cite{mastrolia} the problem of choosing  
suitable master integrals was related to the diagonalization of 
matrices ${\tilde A}$.

Below we present the master integrals and the boundary conditions in the physical region. 
 For all the three families we choose the master 
integrals to be   $f_i = N_0 M^{4 \eps}  \, e^{2 \gamma_{\rm E} \eps} \, g_{i}$,
where $M$ is the overall mass-dimension scaling parameter used to parametrize Mandelstam invariants. 
The 
normalization constant $N_0$ is 
\be
N_0 = 1 + \frac{\pi^2}{6} \ep^2 + \frac{32 \zeta_3 }{3} \ep^3 + \frac{67 \pi^4 \ep^4}{360}.
\ee
Furthermore, to present the master integrals and the results for the limits, we use the following notation 
\be
\begin{split} 
& N_3 = 1 - i\ep\pi - \frac{\pi^2 \ep^2}{6} - \left (\frac{i\pi^3}{6} + 14\zeta_3 \right )\ep^3,
\;\;\;\; 
R_{12}=\sqrt{p_1^4 + (p_2^2 - s)^2 - 2 p_1^2 (p_2^2 + s)},
\\
& R_{13}=\sqrt{p_1^4 + (p_3^2 - t)^2 - 2 p_1^2 (p_3^2 + t)},\;\;\;\; 
R_{34}=\sqrt{p_3^4 + (p_4^2 - s)^2 - 2 p_3^2 (p_4^2 + s)}.
\end{split} 
\ee

In contrast to our previous paper \cite{Henn:2014lfa}, we will not present results for 
all integrals that are needed to construct the non-planar master integrals. 
The reason is that many of these integrals are  the planar ones; they were computed in 
 Ref.~\cite{Henn:2014lfa}.  For the family $N_{34}$, some of the planar  integrals 
need to be re-expressed in new variables, since the parametrization of the family $N_{34}$ 
differs from the parametrization used for   all other families. This is straightforward 
to do, at least in principle. 
Therefore, below we present  our choices of  the genuine non-planar integrals 
and the boundary conditions for them. However, we note that a  complete set of all master 
integrals for the three integral  families can be found in the 
ancillary files. 

Finally, we note that the pictures of master integrals shown 
below are intended to give a general idea of how the 
corresponding master integrals look like, but they  do not show squared  
propagators, numerators and prefactors. Also, in some cases we chose linear combinations 
of integrals as master integrals, and in those cases only one representative figure is given. 

For the family $N_{12}$, there are eight genuine non-planar master integrals. The boundary 
conditions are evaluated  at the point $x \to 0, y \to 1, z \to 1$.

\begingroup
\allowdisplaybreaks
\begin{small}
 \begin{align}
\picturepage{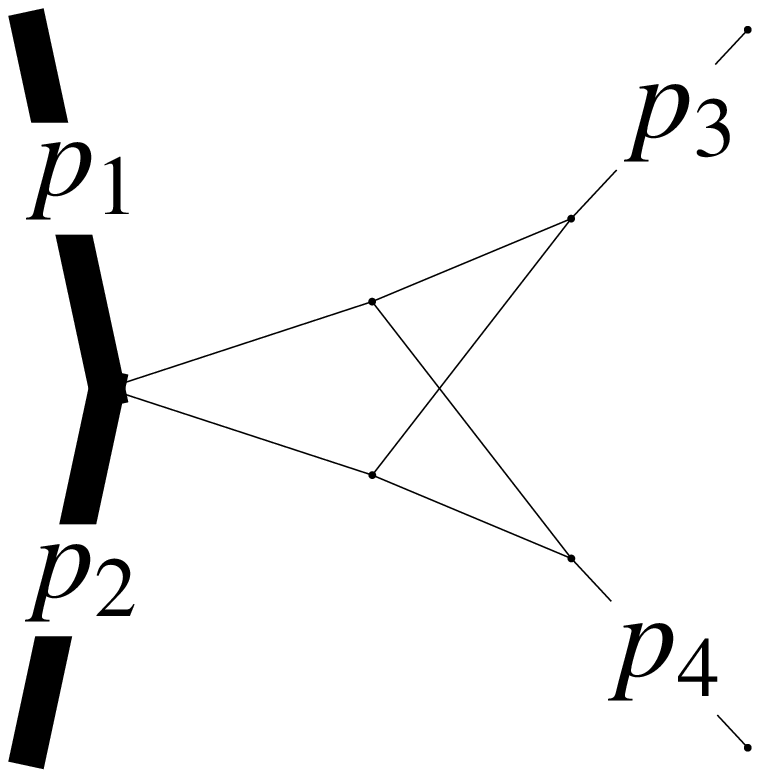}
{  \bea     
  g^{\rm N12}_{28} &=& \eps^4 \; s^2 \;G_{1, 1, 1, 1, 0, 1, 1, 0, 0} 
\,,~~~~~~~~~~~~~~~~~~~~~~~~~~~~~~~~~~~~~~~~~~~~~~~~
  \\
     f^{\rm N12}_{28} &\sim &  e^{2i \pi \ep} \left (
 1 - \frac{5 \pi^2 \ep^2}{6} - 17 \zeta_3 \ep^3  - \frac{17 \pi^4 \ep^4}{36} \right )
\;, \nn 
\eea} \nn \\
\picturepage{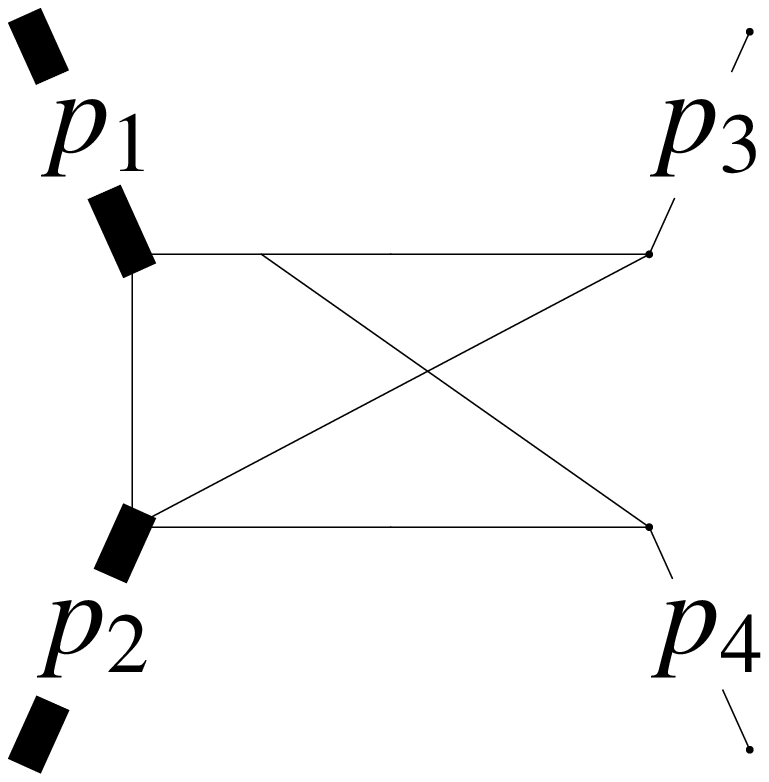}
{  \bea       g^{\rm N12}_{29} &=&  \eps^4 \; p_1^2 s \;  G_{1, 0, 1, 1, 1, 1, 1, 0, 0}
\,,~~~~~~~~~~~~~~~~~~~~~~~~~~~~~~~~~~~~~~~~~~~~~~~~
\\
f^{\rm N12}_{29} &\sim&  
\frac{x^{-4 \ep}}{4}  \Bigg [ 1 + 10 i \pi \ep  - \frac{46  \pi^2 \ep^2 }{3} 
-\left ( 12 i \pi^3 - 16 \zeta_3  \right ) \ep^3 
\nn \\
&+&
    \left ( \frac{386 \pi^4}{45} - 32 i \pi \zeta_3 \right ) \ep^4
 \Bigg ] - i\pi \ep x^{-4\ep} \left [ (z-y)(1-z) \right ]^{-2 \ep}, 
\nn 
\eea} \nn \\
\picturepage{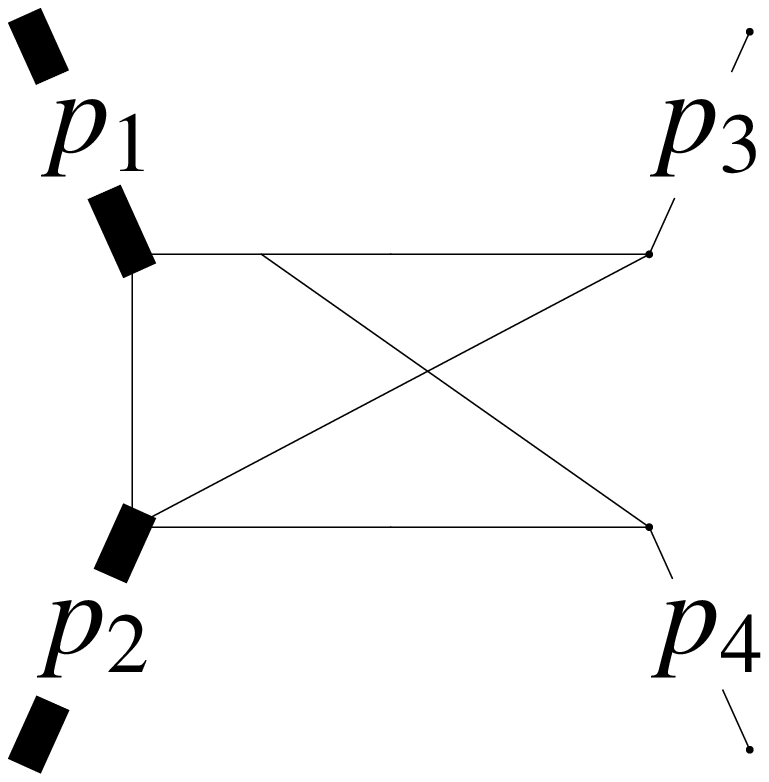}
{  \bea      g^{\rm N12}_{30} &=&   
\eps^4 \left ( (-p_2^2 + t) G_{0, 0, 1, 1, 1, 1, 1, 0, 0} + (u-p_1^2) G_{1, 0, 1, 1, 1, 1, 1, 0, -1} \right )
\,,
 \\
    f^{\rm N12}_{30} &\sim&  
\frac{1}{4} e^{2 i \pi \ep}
-\frac{1}{4} x^{-2\ep} 
+ x^{-4\ep}  \Bigg [ -i \pi \ep + \frac{10 \pi^2 \ep^2}{6} 
\nn \\
& +&   \left ( \frac{4 i \pi^3}{3}  - 2 \zeta_3 \right ) \ep^3
      - \left ( \frac{89 \pi^4}{90} - 4i \pi \zeta_3 \right ) \ep^4 \Bigg ]\;, \nn
\eea} \nn \\
\picturepage{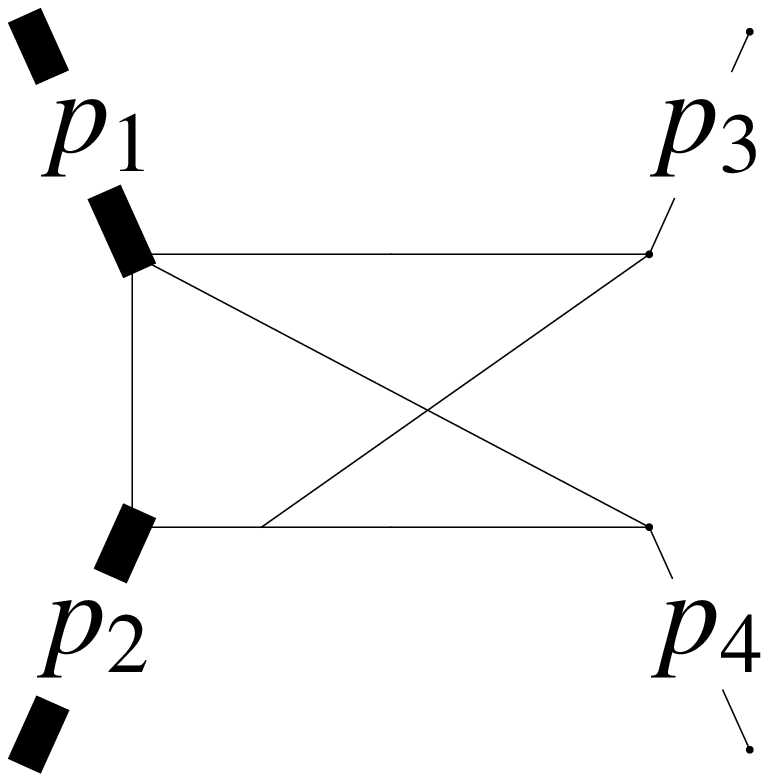}
{  \bea    ~~~~~g^{\rm N12}_{31} &=&  \eps^4 p_2^2 s G_{0, 1, 1, 1, 1, 1, 1, 0, 0}
\,,
 \\
    f^{\rm N12}_{31} &\sim&  1 + 2 i  \pi \ep - \frac{17 \pi^2 \ep^2}{6} 
 -  \left( 3 i \pi^3 + 17 \zeta_3 \right ) \ep^3 + 
  \left ( \frac{67 \pi^4}{36} - 34 i \pi \zeta_3 \right ) \ep^4
\nn
\\
&-&   x^{-2\ep} \left ( 1 - \frac{\pi^2 \ep^2}{6} - 7 \zeta_3 \ep^3 - \frac{\pi^4 \ep^4}{3}   
\right ) 
+ \frac{1}{4} e^{2i\pi \ep} x^{-4 \ep} 
\nn \\
&-& i\pi \ep x^{-4\ep} \left [ (z-y)(1-z) \right ]^{-2 \ep} \;, 
\nn
\eea} \nn \\
\picturepage{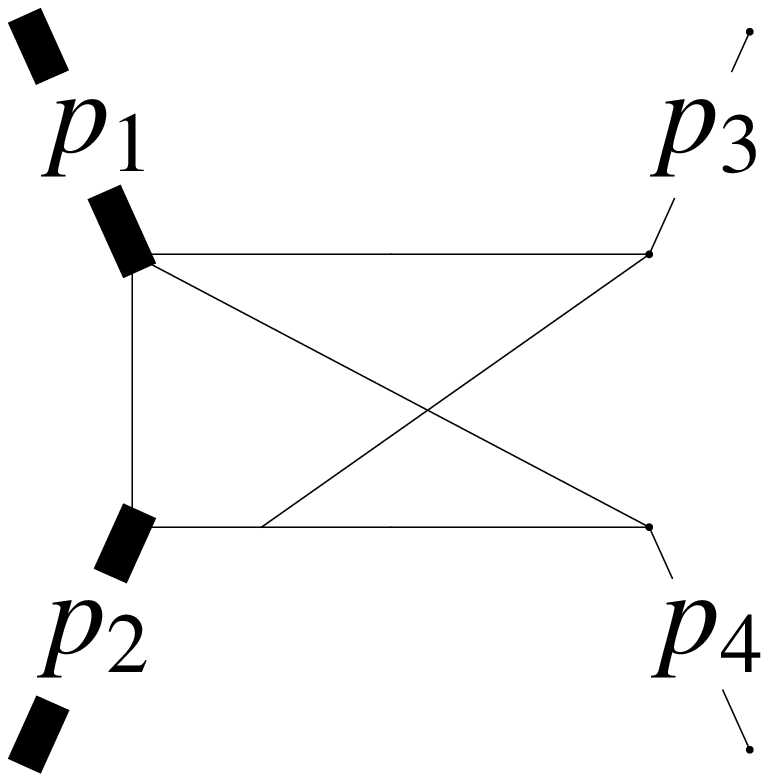}
{  \bea     ~~~~~~~~~ g^{\rm N12}_{32} &=&  \eps^4 \left ( (t-p_1^2) G_{0, 0, 1, 1, 1, 1, 1, 0, 0} 
+ (p_1^2 - s - t) G_{0, 1, 1, 1, 1, 1, 1, 0, -1} \right )
\,,
 \\
    f^{\rm N12}_{32} &\sim& 
-\frac{1}{4} - \frac{i \pi \ep }{2} + \frac{11  \pi^2 \ep^2 }{12} + 
  \left ( \frac{7 i \pi^3}{6} + \frac{17 \zeta_3}{2} \right ) \ep^3 -
 \left ( \frac{55 \pi^4 }{72} - 17 i \pi \zeta_3 \right ) \ep^4 
\nn \\
&+&
x^{-2\ep} \left ( \frac{1}{4} - \frac{\pi^2 \ep^2}{12}  
- \frac{7 \zeta_3 \ep^3}{2}  - \frac{\pi^4 \ep^4}{6} \right )
\;,  \nn
\eea} \nn \\
\picturepage{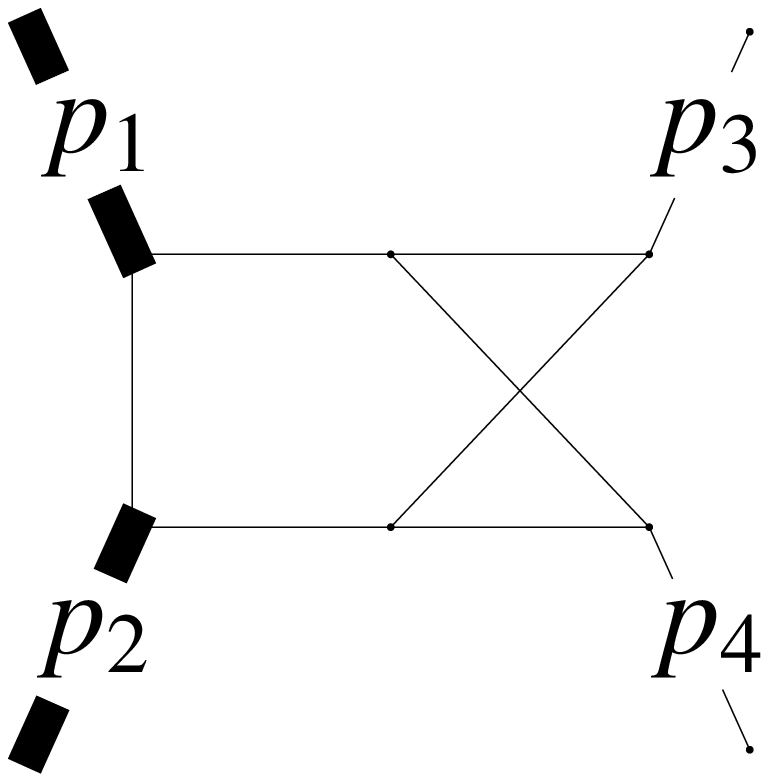}
{  \bea      ~~~~~~~~~~~~~~~~g^{\rm N12}_{33} &=&   
\eps^4 s t \Big [ G_{0, 1, 1, 1, 1, 1, 1, 0, 0} + 
   G_{1, 0, 1, 1, 1, 1, 1, 0, 0}
\nn \\
&-&   G_{1, 1, 1, 1, 1, 1, 1,0, -1} + 
   s G_{1, 1, 1, 1, 1, 1, 1, 0, 0} \Big ] \,,
 \\
    f^{\rm N12}_{33} &\sim&  
-\frac{x^{-2\ep}}{2} \left (1 - \frac{\pi^2 \ep^2}{6} - 7 \zeta_3 \ep^3 - \frac{\pi^4 \ep^4}{3} 
 \right  )
\nn \\
  &+& \frac{x^{-4\ep}}{4} \left ( 
1 + 6 i \pi \ep  - \frac{26 \pi^2 \ep^2 }{3}  + ( 8 \zeta_3  - \frac{20  i  \pi^3}{3} ) \ep^3 
 + \left ( \frac{208  \pi^4  }{45} - 16 i \pi \zeta_3    \right )\ep^4
\right )
\nn \\
& -& i\pi \ep x^{-4\ep} \left [ (z-y)(1-z) \right ]^{-2 \ep} 
, 
\nn
\eea} \nn   \\
\picturepage{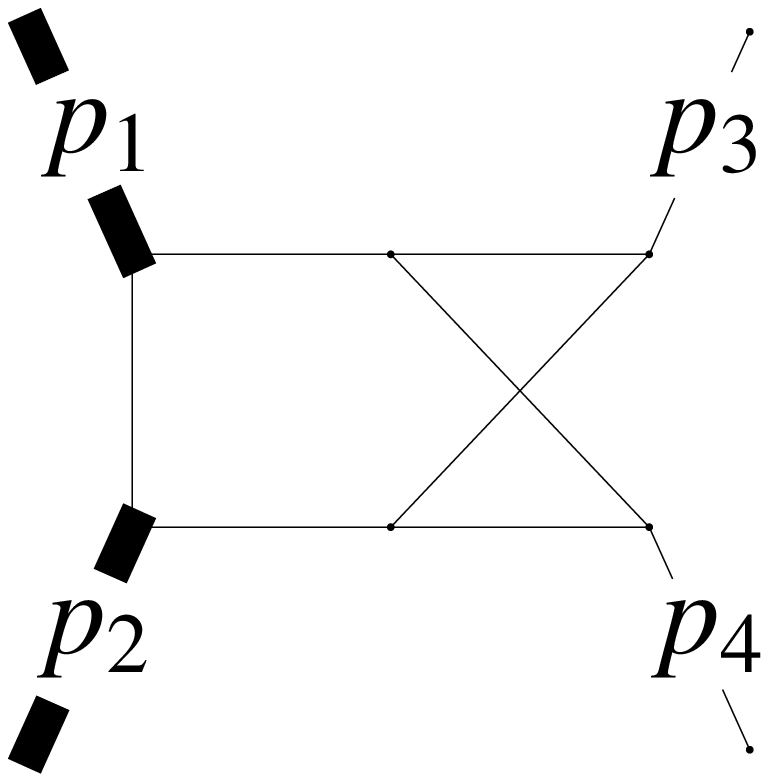}
{  \bea      ~~~~~~~~~~~~~~~~~g^{\rm N12}_{34} &=&-\eps^4 s u G_{1, 1, 1, 1, 1, 1, 1,0,-1}
\,,
 \\
    f^{\rm N12}_{34} &\sim& 
\frac{x^{-2\ep}}{2}\left( 1 - \frac{\pi^2 \ep^2}{6} - 7 \zeta_3 \ep^3 - \frac{\pi^4 \ep^4}{3} \right )
\nn \\
&-& \frac{x^{-4\ep}}{4} \left ( 
1 + 6 i \pi \ep  - \frac{26 \pi^2 \ep^2 }{3}  + ( 8 \zeta_3  - \frac{20  i  \pi^3}{3} ) \ep^3 
 + \left ( \frac{208  \pi^4  }{45} - 16 i \pi \zeta_3    \right )\ep^4
 \right )
\nn \\
&+& i\pi \ep x^{-4\ep} \left [ (z-y)(1-z) \right ]^{-2 \ep} 
\;,
\nn
\eea} \nn \\
\picturepage{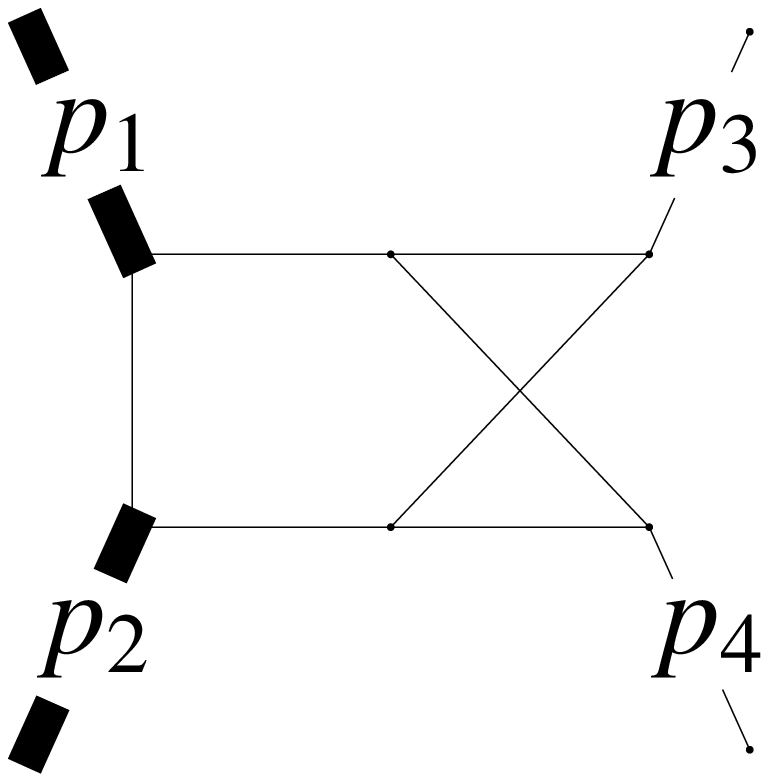}
{  \bea    ~~g^{\rm N12}_{35} &=&  
\eps^4 R_{12}  \Big [ G_{0, 0, 1, 1, 1, 1, 1, 0, 0} -
   G_{0, 1, 1, 1, 1, 1, 1, 0, -1} 
\nn \\
&-& 
   G_{1, 0, 1, 1, 1, 1, 1, 0, -1} 
-
   s G_{1, 1, 1, 1, 1, 1, 1,0,-1} +
   G_{1, 1, 1, 1, 1, 1, 1, 0, -2} \Big ]
\,,
 \\
    f^{\rm N12}_{35} &\sim& 0, \nn
\eea} \nn 
 \end{align}
\end{small}
%
%

There are nine non-planar 
 master integrals in the family $N_{13}$. These integrals, 
together with their limits 
in the kinematic point $x \to 0,~ y \to 1,~ 
z \to 1$  are 
\begin{small}
\begin{align}
\picturepage{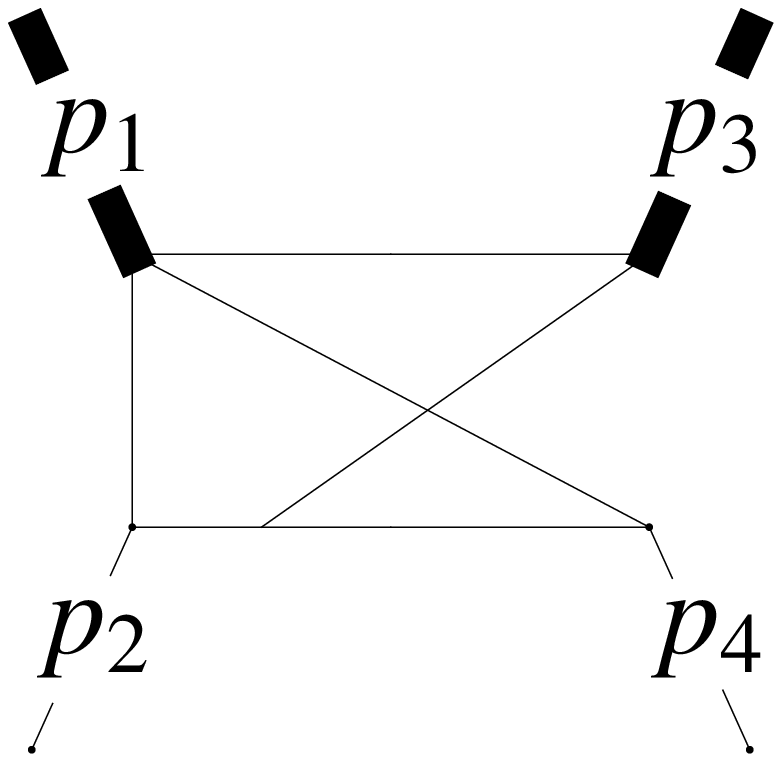}
{  \bea    ~~~~~~g^{\rm N13}_{33} &=& \eps^4 p_3^2 t G_{0, 1, 1, 1, 1, 1, 1, 0, 0}
\,,
\\
    f^{\rm N13}_{33} &\sim&  1 + 2 i \pi \ep 
 - \frac{17 \pi^2 \ep^2}{6} 
-  (3 i \pi^3 + 17 \zeta_3  ) \ep^3 +  
 \left ( \frac{  67 \pi^4}{36} - 34 i \pi  \zeta_3  \right ) \ep^4 
\nn \\
&-&x^{-2\ep} \Big (1 - \frac{ \pi^2 \ep^2 }{6} -  7 \zeta_3 \ep^3  - \frac{\pi^4 \ep^4 }{3}  \Big )
+\frac{e^{2i\pi \ep}}{4}  x^{-4\ep}
\nn \\
&-& i \pi \ep x^{-4\ep} \left [ (z-y)(1-z) \right ]^{-2\ep}
,\;\;\;   \nn
\eea} \nn \\
\picturepage{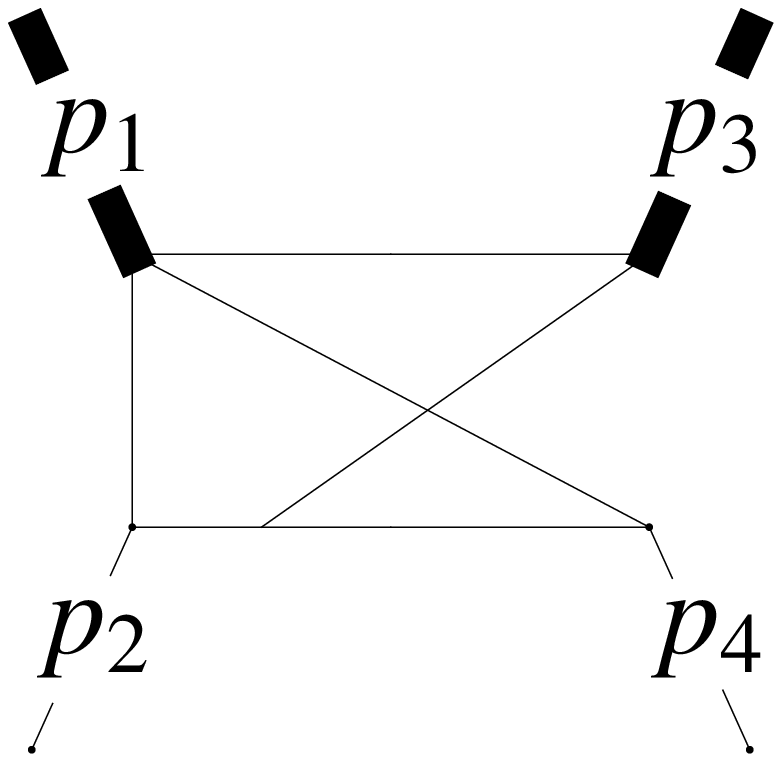}
{  \bea    ~~~~~~g^{\rm N13}_{34} &=&  
\eps^4 ( (t-p_1^2 + p_3^2 ) 
G_{0, 0, 1, 1, 1, 1, 1, 0, 0} 
+  (u-p_3^2) G_{0, 1, 1, 1, 1, 1, 1, 0, -1} )
\,,
 \\
    f^{\rm N13}_{34} &\sim& 
\frac{1}{2} + i  \pi \ep  - \frac{3  \pi^2 \ep^2 }{2} 
-  \left (\frac{5 i \pi^3}{3} + 9 \zeta_3 \right ) \ep^3  
 +   \left (  \frac{193 \pi^4}{180} - 18 i \pi  \zeta_3  \right ) \ep^4
\nn \\ 
& -&  x^{-2\ep}  \left ( \frac{3}{4}  - \frac{ \pi^2 \ep^2}{4} - \frac{21 \zeta_3 \ep^3 }{2} 
 - \frac{\pi^4 \ep^4}{2} 
      \right )
\nn \\
& +&  x^{-4\ep} \left ( \frac{1}{4} + \frac{i \pi \ep}{2} - \frac{\pi^2 \ep^2}{2} - 
    \frac{i \pi^3 \ep^3}{3} + \frac{\pi^4 \ep^4}{6} \right )
\nn
\\
&-&i \pi \ep x^{-4\ep} \left [ (z-y)(1-z) \right ]^{-2\ep}
, \;\;\;\  \nn
\eea} \nn  
\\
\picturepage{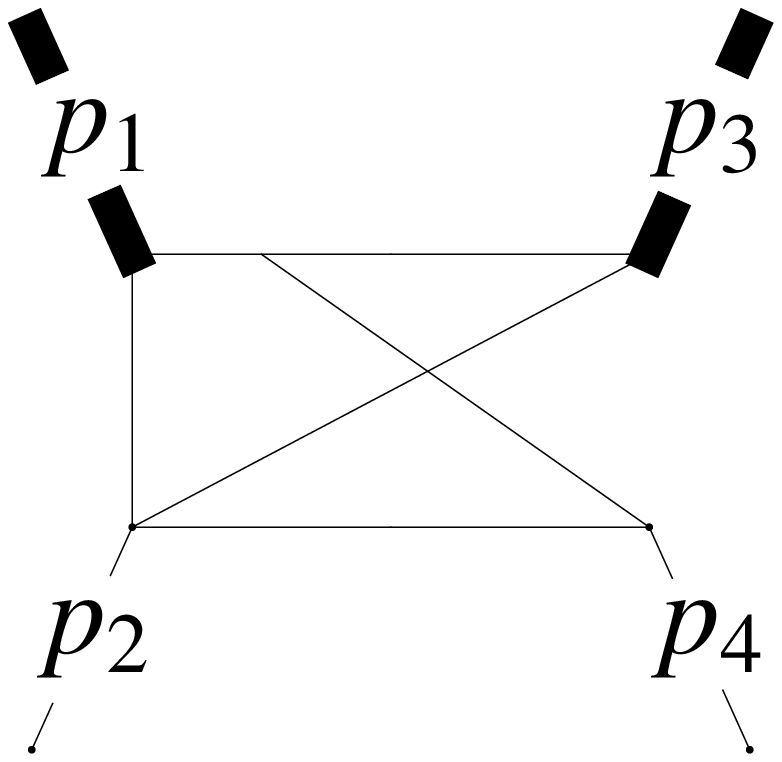}
{  \bea      ~~~~~~~~~~g^{\rm N13}_{35} &=&  
\eps^4 \left ( p_1^2 (p_3^2 - s) + p_3^2 (-p_3^2 + s + t) \right ) G_{1, 0, 1, 1, 1, 1, 1, 0, 0}\,,
\\
    f^{\rm N13}_{35} &\sim& 
  -x^{-2\ep} \left ( \frac{1}{2} - \frac{ \pi^2 \ep^2}{12}  
     - \frac{7}{2} \zeta_3 \ep^3  
- \frac{ \pi^4 \ep^4}{6}
\right )  
\nn \\
  & +&  x^{-3\ep} \left (1 + i \ep \pi  - \frac{2 \pi^2 \ep^2}{3} -
      \left (\frac{i \pi^3}{3} - 2 \zeta_3 \right ) \ep^3 + 
      \left ( \frac{\pi^4}{10} + 2 i \pi \zeta_3 \right )  \ep^4 \right )
\nn \\
& -&  x^{-4\ep} \left ( \frac{1}{2} + \frac{2  i \pi \ep}{3} - \frac{5  \pi^2 \ep^2}{12} 
     -  \left ( \frac{5 i \pi^3}{18} + \frac{\zeta_3}{2} \right ) \ep^3  + 
      \left ( \frac{53 \pi^4}{360} - \frac{i \pi \zeta_3}{3} \right ) \ep^4  \right ) 
\nn \\
&-& \frac{i \pi \ep}{3} x^{-4 \ep} \left [ (z-y)(1-z) \right ]^{-3 \ep}N_3 
,\nn
\eea} \nn 
\\
\picturepage{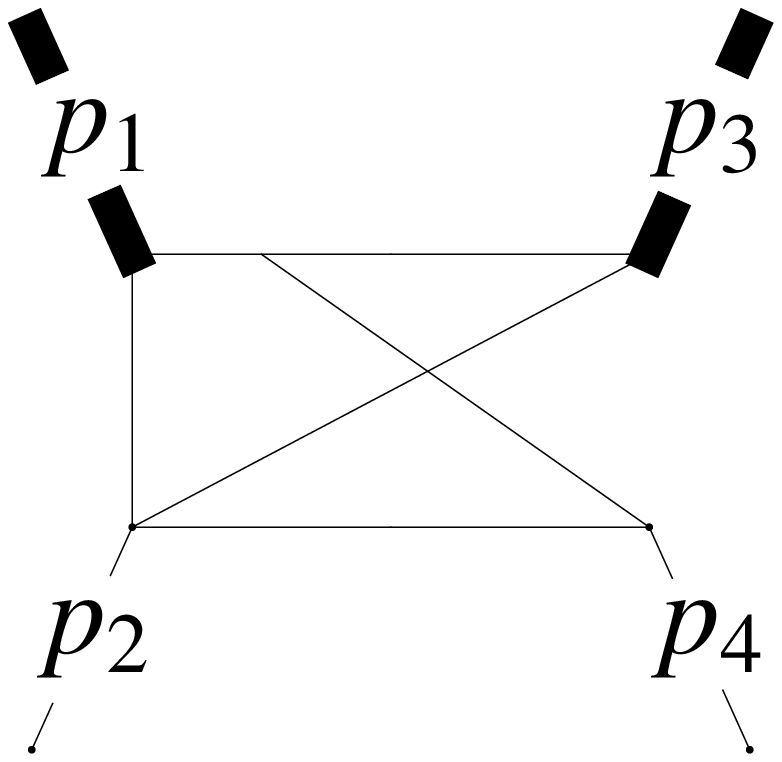}
{  \bea      g^{\rm N13}_{36} &=&  
\eps^4 R_{13} \Big (  G_{1, -1, 1, 1, 1, 1, 1, 0, 0} - 
   G_{1, 0, 1, 1, 1, 1, 1, 0, -1} 
\nn \\
& +&  (s-p_3^2) G_{1, 0, 1, 1, 1, 1, 1, 0, 0} \Big  )
\,,~~~~~~~~~~~~~~~~~~~~~~~~~~~~~~~~~~~~~~~~~~~~~~
 \\
    f^{\rm N13}_{36} &\sim& 0 ,\nn
\eea} \nn 
\\
\picturepage{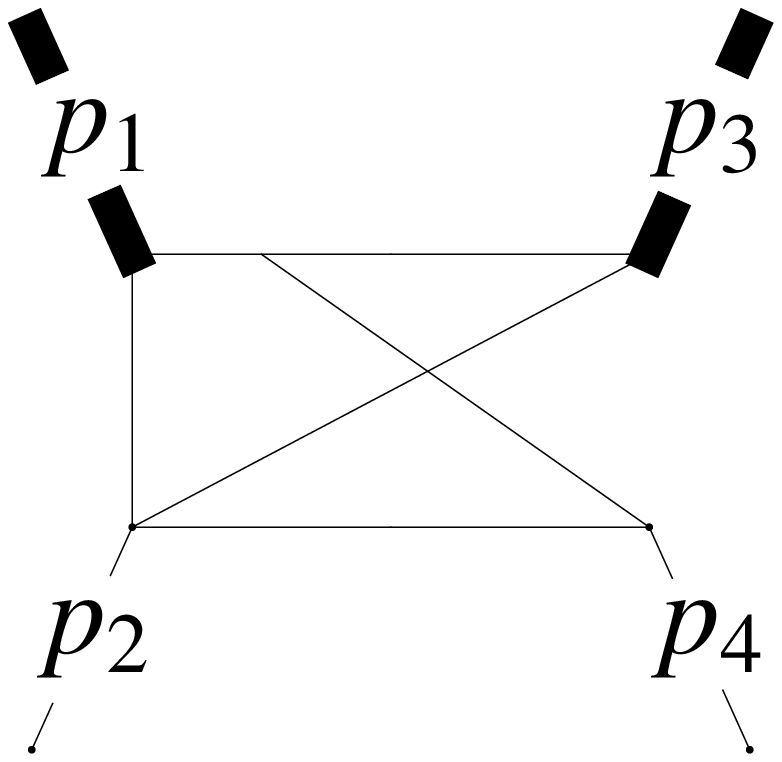}
{  \bea      ~~~~g^{\rm N13}_{37} &=&  
\eps^4 \left ( t G_{1, 0, 0, 1, 1, 1, 1, 0, 0} 
     + (-p_3^2 + s) G_{1, 0, 1, 1, 1, 1, 1, -1, 0} \right )
\,,
 \\
    f^{\rm N13}_{37} &\sim& 
-x^{-4 \ep} \Big  [ 
\frac{1}{4} 
+\frac{i \pi \ep}{6}
+\frac{\pi ^2 \ep^2}{12}
-\left(\frac{\zeta_3}{2}-\frac{i \pi ^3}{18}\right) \ep^3
- \left (  \frac{7 \pi ^4}{360} + \frac{i \pi  \zeta_3}{3}\right) \ep^4
\Big ]
\nn \\
&+& x^{-3 \ep} \Big [  
\frac{1}{2} 
+\frac{i \pi \ep}{2}
-\frac{\pi ^2 \ep^2}{3}
+ \left(\zeta_3-\frac{i \pi ^3}{6}\right) \ep^3
+\left(\frac{\pi ^4}{20}+i \pi  \zeta_3 \right) \ep^4 
\Big ]
\nn \\
&-& x^{-2 \ep} \Big [
\frac{1}{4} 
-\frac{\pi ^2 \ep^2}{12}
-\frac{7  \zeta_3 \ep^3 }{2}
- \frac{\pi ^4 \ep^4}{6} \Big ]
\nn \\
&-& \frac{i \ep \pi}{3}   x^{-4 \ep} \left [ (z-y) (1-z) \right ]^{-3 \ep} N_3
 ,
 \nn
\eea} 
\nn 
\\
\picturepage{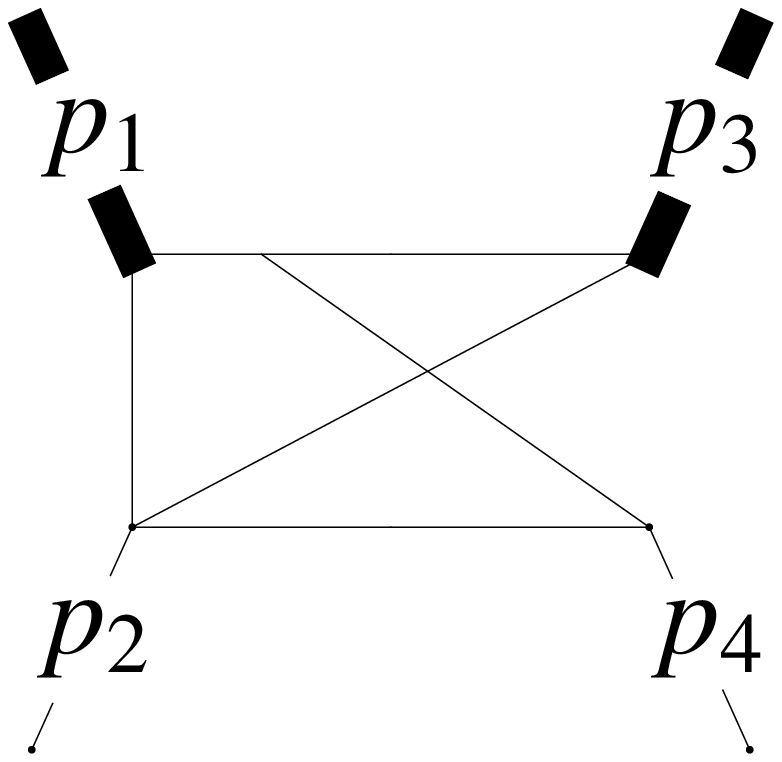}
{  \bea    ~~~~~~~~~~~~~~f^{\rm N13}_{38} &=&  
\eps^4 \left ( -t G_{0, 0, 1, 1, 1, 1, 1, 0, 0} 
       + (-p_3^2 + s + t) G_{1, 0, 1, 1, 1, 1, 1, 0, -1}
\right )\;,
 \\
    f^{\rm N13}_{38} &\sim& 
\frac{\pi ^2 \ep^2}{12}
+\ep^3 \left(\frac{\zeta_3}{2}+\frac{i \pi
   ^3}{6}\right) - \ep^4 \left(\frac{17 \pi ^4}{120}-i \pi  \zeta_3\right)
\nn \\
&+& x^{-2 \ep} \left[
\frac{1}{2}
-\frac{\pi
   ^2 \ep^2}{12}
-\frac{7  \zeta_3 \ep^3 }{2} 
 -\frac{\pi ^4 \ep^4}{6} 
\right ]
\nn \\
&-& x^{-3 \ep} \left[ 
1
+i\pi  \ep -\frac{2 \pi ^2 \ep^2}{3}
+ \left(2 \zeta_3 - \frac{i \pi ^3}{3}\right) \ep^3
+ \left( \frac{\pi^4}{10} + 2 i \pi  \zeta_3\right) \ep^4
\right] 
\nn \\
&+& x^{-4 \ep} \left[ 
\frac{1}{2} +\frac{2 i \pi  \ep}{3} 
-\frac{5 \pi ^2\ep^2}{12}
-  \left( \frac{\zeta_3}{2}+\frac{5 i \pi ^3}{18}\right) \ep^3
+ \left(\frac{53 \pi ^4}{360}-\frac{i \pi 
   \zeta_3}{3}\right) \ep^4  
\right]
\nn \\
&+& \frac{i \ep \pi}{3}   x^{-4 \ep} \left [ (z-y) (1-z) \right ]^{-3 \ep} N_3\;,
\nn \eea} 
\nn  \\
\picturepage{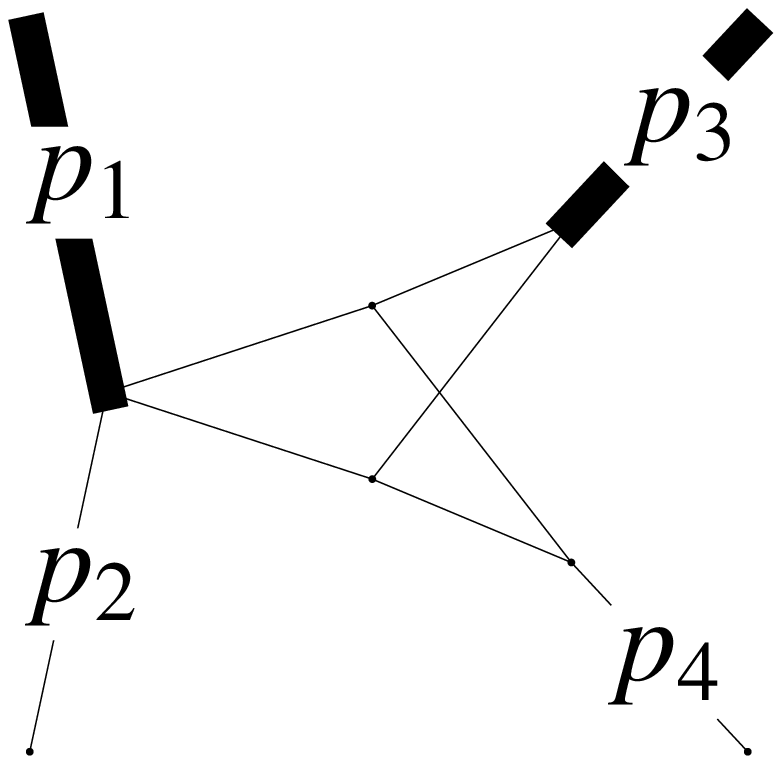}
{  \bea     ~~~~~~~~g^{\rm N13}_{40} &=&  
\eps^4 (p_3^2 - s)^2 G_{1, 1, 1, 1, 0, 1, 1, 0, 0}
\,,
 \\
    f^{\rm N13}_{40} &\sim&  
1+2 i \pi  \ep -\frac{17
   \pi ^2 \ep^2}{6} 
-\ep^3 \left(17 \zeta_3 + 3 i \pi ^3\right)
+ \ep^4 \left ( \frac{67 \pi ^4}{36}-34 i \pi  \zeta_3\right)
\nn \\
&-& x^{-\ep}  \left[  
2
+2 i \pi  \ep
-2 \pi ^2 \ep^2
- \left(12 \zeta_3 +\frac{4 i \pi ^3}{3}\right) \ep^3
+ \left(\frac{\pi ^4}{30} - 12 i \pi  \zeta_3\right) \ep^4
\right ]
\nn \\
&+& x^{-2\ep} \left [ 1 -\frac{\pi ^2 \ep^2}{6}  -7 \ep^3 \zeta_3 
 -\frac{ \pi ^4 \ep^4 }{3} 
\right ],
\nn
\eea} \nn \\
\picturepage{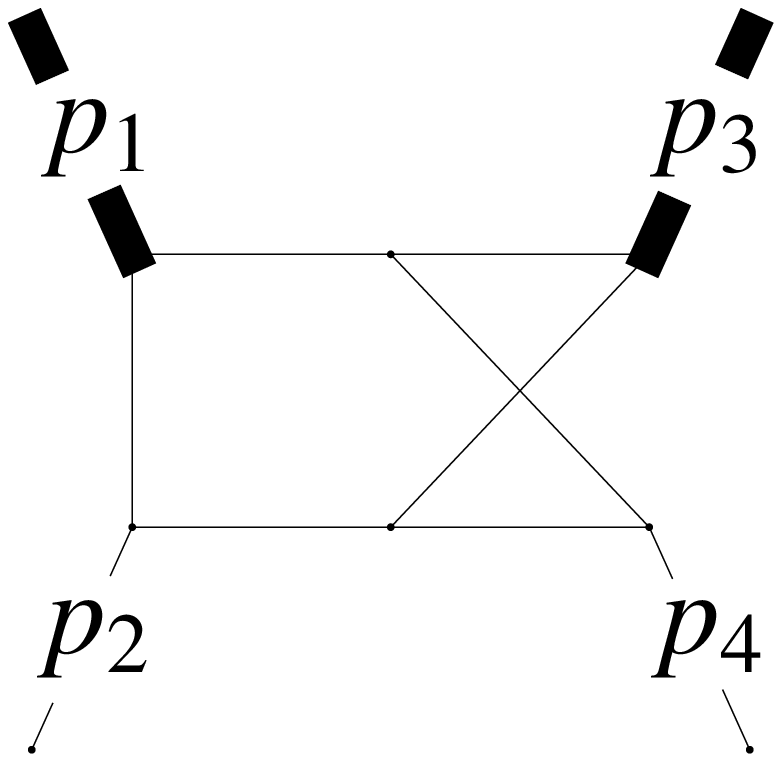}
{  \bea     ~~~~~~~~~~~~~~~g^{\rm N13}_{42} &=&
\eps^4 s t \Big [ G_{0, 1, 1, 1, 1, 1, 1, 0, 0} + 
   G_{1, 0, 1, 1, 1, 1, 1, 0, 0} - 
   G_{1, 1, 1, 1, 1, 1, 1, 0, -1} 
\\
 & -&  
   p_3^2 G_{1, 1, 1, 1, 1, 1, 1, 0, 0} + 
   s G_{1, 1, 1, 1, 1, 1, 1, 0, 0} \Big ]
  \,,
 \nn \\
    f^{\rm N13}_{42} &\sim& 
x^{-4 \ep} \left(
\frac{3}{4} +\frac{7 i \pi 
   \ep}{6}
-\frac{11 \pi ^2 \ep^2}{12}
- \left ( \frac{\zeta_3}{2} + \frac{11}{18} i \pi ^3 \right ) \ep^3
- \left ( \frac{1}{3} i \pi  \zeta_3+\frac{113 \pi ^4}{360} \right ) \ep^4
\right )
\nn \\
&-&x^{-3 \ep} \left(
1
+i \pi  \ep
-\frac{2 \pi ^2 \ep^2}{3}
+ \left ( 2 \zeta_3-\frac{1}{3} i \pi ^3 \right ) \ep^3
+ \left ( 2 i \pi   \zeta_3 + \frac{\pi ^4 }{10} \right ) \ep^4
\right)
\nn \\
&-&x^{-2\ep} \left(
\frac{3}{2}
-\frac{\pi ^2\ep^2}{4}
-\frac{21 \ep^3 \zeta_3}{2}
-\frac{\pi ^4 \ep^4}{2}
\right)
\nn \\
&+&x^{-\ep} \left(
2 + 2 i \pi  \ep - 2 \pi ^2 \ep^2
- \left ( 12 \zeta_3+\frac{4}{3} i \pi ^3 \right ) \ep^3
- \left ( 12 i \pi   \zeta_3-\frac{\pi ^4}{30} \right ) \ep^4
\right)
\nn \\
&+& \frac{i \pi \ep }{3} \left[ (z-y)(1-z) \right 
]^{-3 \ep} x^{-4 \ep} N_3 
- 2 i \pi  \ep [(z-y)(1-z)]^{-2 \ep}  x^{-3 \ep}
, \nn
\eea} \nn 
\\
\picturepage{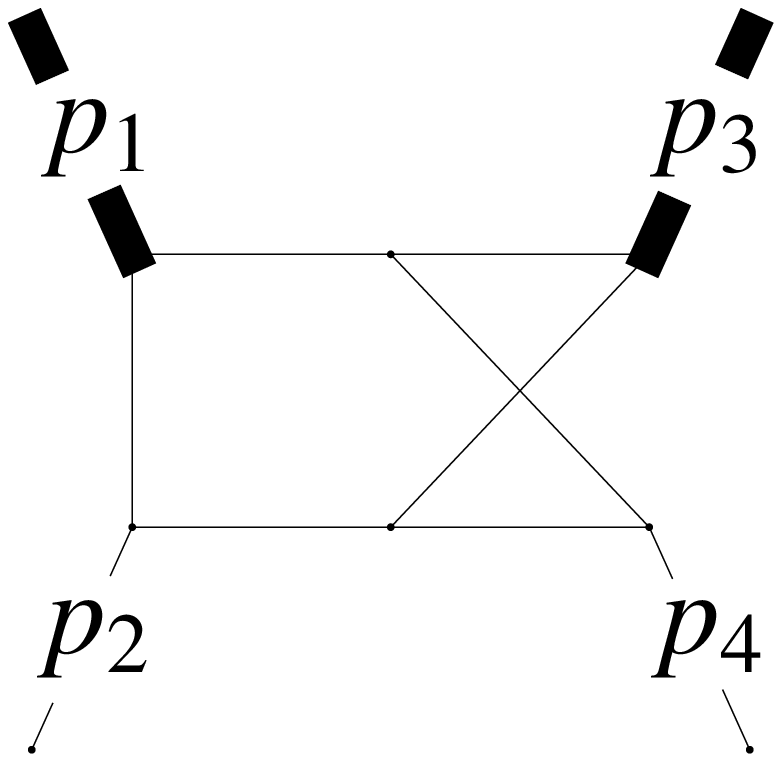}
{  \bea      g^{\rm N13}_{43} &=&  
\eps^4 \left [ p_1^2 (p_3^2 - s) + s ( s + t -p_3^2 ) \right ]
 G_{1, 1, 1, 1, 1, 1, 1, 0, -1},
\;~~~~~~~~~~~~~~~~
 \\
    f^{\rm N13}_{43} &\sim& 
i \pi  \ep \left [(z-y)(1-z) \right ]^{-2 \ep} 
\left(6 x^{-3 \ep}-3 x^{-4 \ep}\right) \nn
\\
&-&
2 i \pi  \ep \left [(z-y)(1-z) \right ]^{-3\ep} x^{-4 \ep} N_3\;, 
\nn
\eea} \nn 
\end{align}
\end{small}

Non-planar  master integrals that appear for the family $N_{34}$ are shown below.
The boundary conditions are derived by considering the limit $x \to 0$, $y \to 0$ and $z \to 1$. The results 
read
\begin{small}
\begin{align}
\picturepage{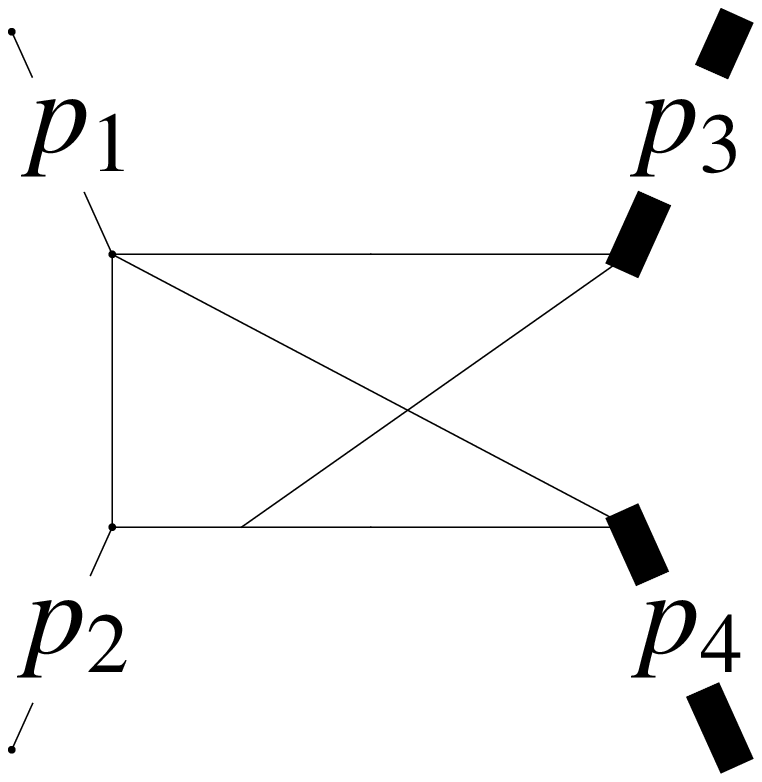}
{  \bea     
~~~~~~~g^{\rm N34}_{37} &=& 
\eps^4 (p_3^2 (p_4^2 - t) + p_4^2 (s+t -p_4^2 )) G_{0, 1, 1, 1, 1, 1, 1, 0, 0},
\,
  \\
f^{\rm N34}_{37} &\sim & 
-x^{-4 \ep} \left [ 
\frac{1}{2} +\frac{2 i \pi \ep}{3} -\frac{5 \pi ^2 \ep^2}{12} 
-\ep^3\left(\frac{\zeta_3}{2}+\frac{5 i \pi ^3}{18}\right)
-\ep^4 \left(-\frac{53 \pi ^4}{360}+\frac{i \pi  \zeta_3}{3}\right)
\right ]
\nn \\
&+&x^{-3 \ep} \left [ 
1 +i \pi \ep -\frac{2 \pi ^2 \ep^2}{3}
+\ep^3 \left(2 \zeta_3-\frac{i \pi ^3}{3}\right)
+ \ep^4 \left(\frac{\pi ^4}{10} +2 i \pi  \zeta_{3}\right)
\right]
\nn \\
&-& x^{-2 \ep} \left[
\frac{1}{2} -\frac{\pi^2 \ep^2}{12} -\frac{7 \ep^3 \zeta_3}{2} - \frac{\pi ^4 \ep^4}{6}\right ]
\nn \\
&-&\frac{i \pi \ep }{3}  \left [ 4 y^2 (1-z) \right ]^{-3 \ep} x^{-4 \ep}N_3,  \nn 
\eea} \nn \\
\picturepage{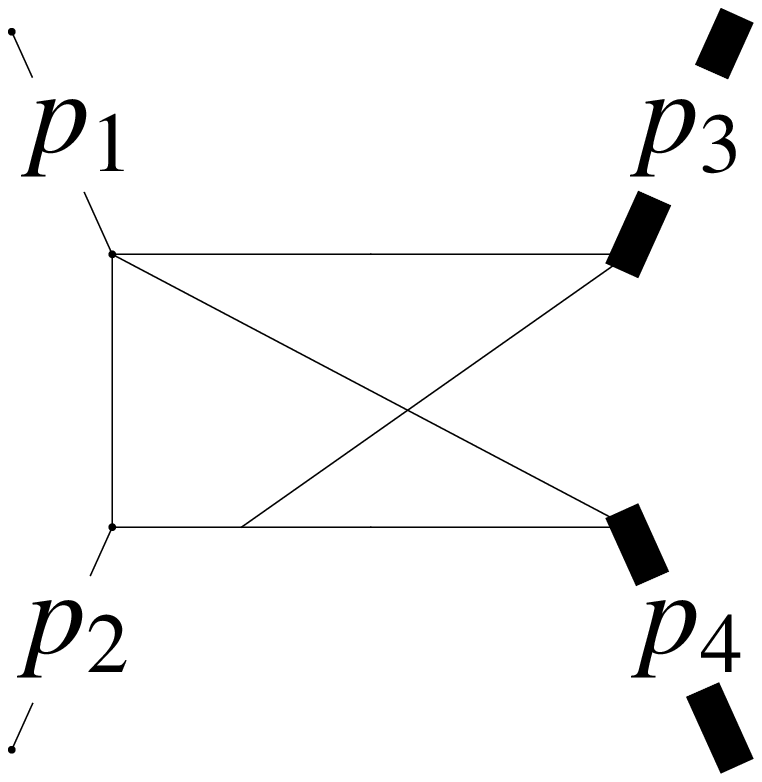}
{  \bea     
  g^{\rm N34}_{38} &=& \eps^4 \Big[  ( p_4^2 - t) G_{-1, 1, 1, 1, 1, 1, 1, 0, 0} 
   - (p_4^2 - t) G_{0, 1, 1, 1, 1, 1, 1, 0, -1} 
\nn \\
   &-&  (p_3^2 p_4^2 - p_3^2 t + s t) 
 G_{0, 1, 1, 1, 1, 1, 1, 0, 0} \Big ]
\,,~~~~~~~~~~~~~~~~~~~~~~~~~~~~~
  \\
     f^{\rm N34}_{38} &\sim &  
-x^{-3 \ep} \left [ 
\frac{1}{2} + \frac{i \pi  \ep}{2} - \frac{\pi ^2 \ep^2}{3}
 + \ep^3 \left( \zeta_3 -\frac{i
   \pi ^3}{6}\right)
+ \ep^4 \left( \frac{\pi ^4}{20}
+ i \pi  \zeta_3\right)
\right ]
\nn \\
&+& x^{-2 \ep}
   \left [ \frac{1}{4} -\frac{\pi ^2
   \ep^2}{12} -\frac{7 \ep^3 \zeta_3}{2}
-\frac{\pi ^4 \ep^4 }{6} 
\right ]
\nn \\
&+& x^{-4 \ep} \left [ 
\frac{1}{4} +\frac{i \pi  \ep}{2} -\frac{\pi^2\ep^2}{2}
-\frac{1}{3} i \pi ^3 \ep^3 + \frac{\pi ^4 \ep^4}{6}
\right ] 
,  \nn 
\eea} \nn \\
\picturepage{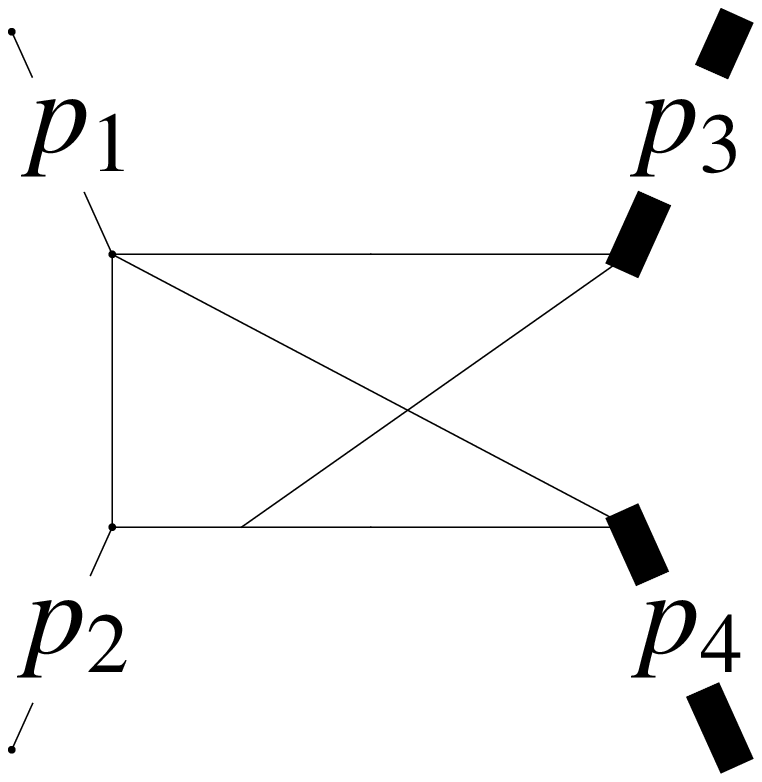}
{  \bea     
  ~~~~~~~~~~g^{\rm N34}_{39} &=&
\eps^4 \Big [ -s G_{0, 1, 1, 1, 1, 1, 0, 0, 0} 
      + (p_4^2 - s - t) (-G_{0, 0, 1, 1, 1, 1, 1, 0, 0} 
\nn \\
&+& 
      G_{0, 1, 1, 0, 1, 1, 1, 0, 0} 
    -
      G_{0, 1, 1, 1, 1, 0, 1, 0, 0} 
\nn \\
&-&  G_{0, 1, 1, 1, 1, 1, 0, 0, 0} + 
      G_{0, 1, 1, 1, 1, 1, 1, 0, -1} + 
      p_4^2 G_{0, 1, 1, 1, 1, 1, 1, 0, 0} )\Big ]
\,,
  \\
     f^{\rm N34}_{39} &\sim &  
\frac{\pi ^2 \ep^2}{12}
+\ep^3 \left(\frac{\zeta_3}{2}+\frac{i \pi
   ^3}{6}\right)
+\ep^4 \left(-\frac{17 \pi ^4}{120}+i \pi  \zeta_3\right)
\nn \\
&+& x^{-4 \ep} \left(
+\frac{1}{2}
+\frac{2 i \pi  \ep}{3}
-\frac{5 \pi ^2
   \ep^2}{12}
+\ep^3 \left(-\frac{\zeta_3}{2}-\frac{5 i \pi ^3}{18}\right)
+\ep^4 \left(\frac{53 \pi ^4}{360}-\frac{i \pi 
   \zeta_3}{3}\right)
\right)
\nn \\
&-&x^{-3 \ep} \left(
1
+i \pi  \ep
-\frac{2 \pi ^2 \ep^2}{3}
+\ep^3 \left(2 \zeta_3-\frac{i \pi ^3}{3}\right)
+\ep^4 \left(
\frac{\pi^4}{10} + 2 i \pi  \zeta_3\right)
\right)
\nn \\
&+&x^{-2 \ep} \left(
\frac{1}{2}
-\frac{\pi ^2 \ep^2}{12}
-\frac{7 \ep^3\zeta_3}{2}
-\frac{1}{6} \pi ^4 \ep^4
\right)
\nn \\
&+&\frac{i \pi}{3}   \ep \left [ 4 y^2 (1-z) \right ]^{-3 \ep} 
x^{-4 \ep} N_3, \nn 
\eea} \nn \\ 
\picturepage{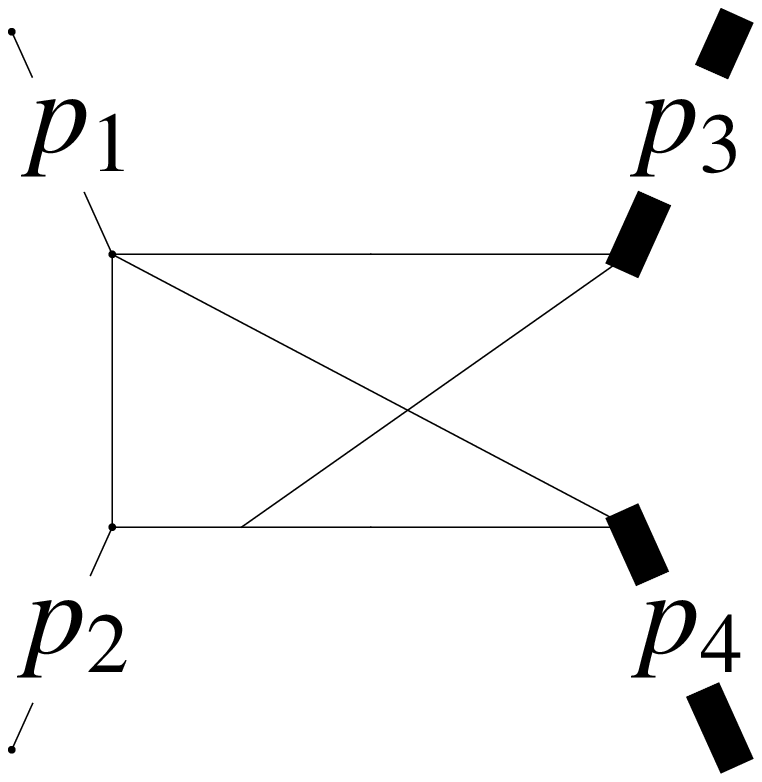}
{  \bea     
  g^{\rm N34}_{40} &=& 
\eps^4 R_{34} \Big [  G_{0, 0, 1, 1, 1, 1, 1, 0, 0} 
   - G_{0, 1, 1, 0, 1, 1, 1, 0, 0} 
\nn \\
 &+& 
   G_{0, 1, 1, 1, 1, 0, 1, 0, 0} 
   + G_{0, 1, 1, 1, 1, 1, 0, 0, 0} 
-  
   G_{0, 1, 1, 1, 1, 1, 1, -1, 0} 
\nn \\
&-& 
   G_{0, 1, 1, 1, 1, 1, 1, 0, -1} - 
   t G_{0, 1, 1, 1, 1, 1, 1, 0, 0} \Big ]
\,,~~~~~~~~~~~~~~~~~~~~~~~~~~~~~~
  \\
     f^{\rm N34}_{40} &\sim &  0,  \nn 
\eea} \nn \\
\picturepage{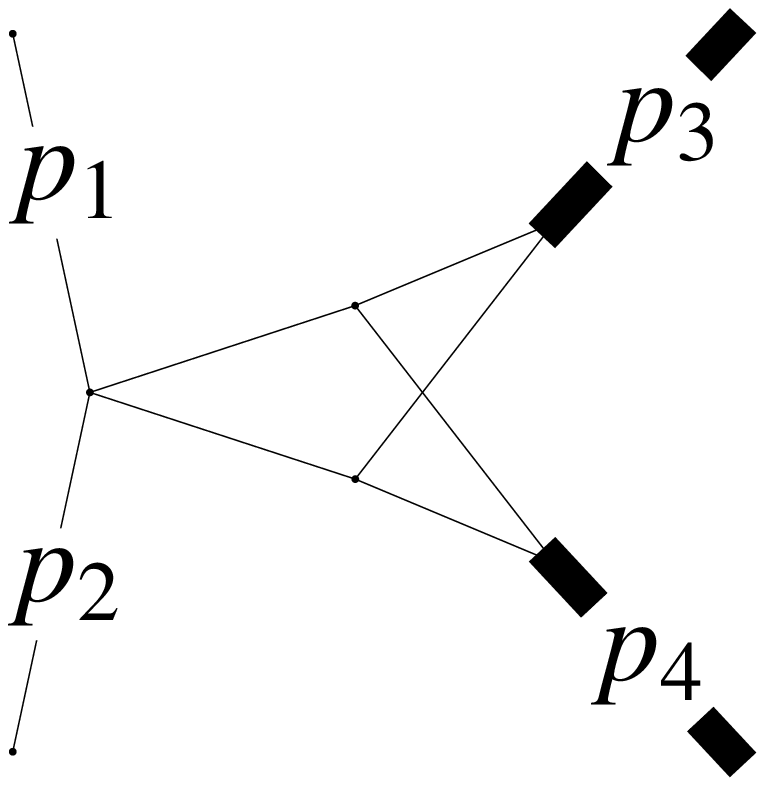}
{  \bea     
  g^{\rm N34}_{47} &=& 
\eps^4 (p_3^4 + (p_4^2 - s)^2 - 2 p_3^2 (p_4^2 + s)) G_{1, 1, 1, 1, 0, 1, 1, 0, 0}
\,,~~~~~~~~~~~~~~~~~~~~~~~
  \\
     f^{\rm N34}_{47} &\sim & 0 
\nn 
\eea} \nn \\
\picturepage{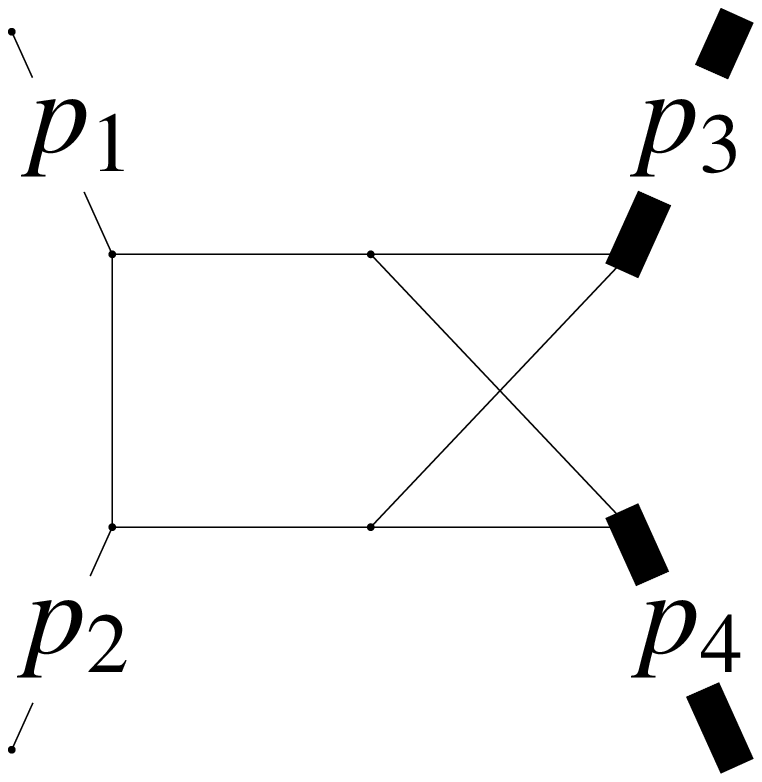}
{  \bea     
  ~~~~~~~g^{\rm N34}_{50} &=& \left [ 2 (p_3^2 + p_4^2 - 
   s) \right ]^{-1} \eps^4 s 
  \Big [ -2 p_3^2 p_4^2 G_{0, 1, 1, 1, 1, 1, 1, 0, 0} - 
   2 p_3^2 p_4^2 G_{1, 0, 1, 1, 1, 1, 1, 0,0} \nn
\\
 &-& (p_3^4 + (p_4^2 - s) (p_4^2 - s - t) - p_3^2 (2 s + t)) 
     G_{1, 1, 1, 1, 1, 1, 1, 0, -1} \nn
\\
& +& (2 p_3^2 p_4^2 - p_3^2 t - p_4^2 t + 
      s t) ( G_{0, 1, 1, 1, 1, 1, 1, 0, 0} 
+ 
      G_{1, 0, 1, 1, 1, 1, 1, 0, 0} 
\\
&-& 
      G_{1, 1, 1, 1, 1, 1, 1, 0, -1}
-
      ( p_3^2 +  
      p_4^2  - 
      s ) G_{1, 1, 1, 1, 1, 1, 1, 0, 0}  ) \Big ]
\,,~~~ \nn
  \\
     f^{\rm N34}_{50} &\sim &  
-x^{-4 \ep} \left(
\frac{1}{2} + \frac{2 i \pi 
   \ep}{3} -\frac{5 \pi ^2 \ep^2}{12}
 -\left(\frac{\zeta_3}{2}+\frac{5 i \pi ^3}{18}\right) \ep^3 
+  \left(+\frac{53 \pi ^4}{360}-\frac{i \pi  \zeta_3}{3}\right) \ep^4 
\right) \nn
\\
&+&x^{-3 \ep} \left(
1 +i \pi \ep -\frac{2 \pi ^2 \ep^2}{3}
+ \left(2 \zeta_3-\frac{i \pi ^3}{3}\right) \ep^3 
+ \left(\frac{\pi ^4}{10}+2 i \pi  \zeta_3 \right) \ep^4 
\right) \nn
\\
&-&x^{-2 \ep} \left(
\frac{1}{2}
-\frac{\pi^2\ep^2}{12}
-\frac{7 \ep^3 \zeta_3}{2}
-\frac{\pi^4 \ep^4}{6}
\right) \nn
\\
&+&\frac{5 i \pi \ep  }{3}  \left [4 y^2 (1-z)   \right ]^{-3 \ep} x^{-4 \ep} N_3
\;,
 \nn 
\eea} \nn \\
\picturepage{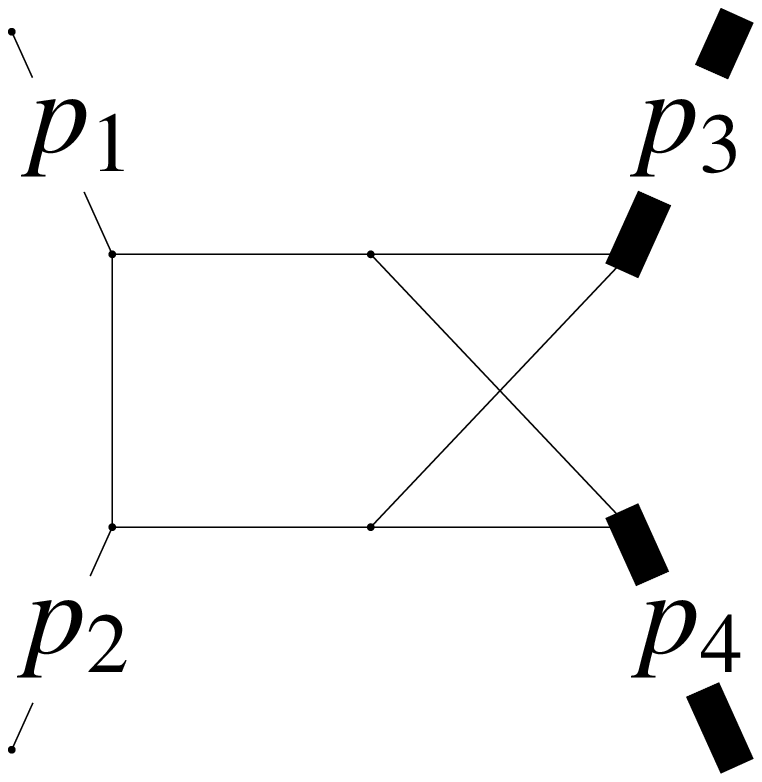}
{  \bea     
  g^{\rm N34}_{51} &=& 
-\left [ p_3^2 + p_4^2 - s \right ]^{-1}
 \eps^4 R_{34}  \Big [  (t-p_4^2) ( s - p_3^2 - p_4^2) G_{0, 1, 1, 1, 1, 1, 1, 0, 0} 
\nn \\
&+& (t-p_3^2)(s-p_3^2-p_4^2)G_{1, 0, 1, 1, 1, 1, 1, 0, 0} 
\nn \\
&-& (p_3^2 + p_4^2 - 
       s) s (G_{1, 1, 1, 1, 1, 1, 1, 0, -1} + 
       t G_{1, 1, 1, 1, 1, 1, 1, 0, 0} ) \Big ] 
\,,~~~~~~~~~~~~~~~~~
  \\
     f^{\rm N34}_{51} &\sim &  0, \nn 
\eea} \nn 
\end{align} 
\end{small}

\endgroup

To illustrate how analytic results look like, we provide contributions through ${\cal O}(\ep^2)$
for three different master integrals.  We introduce the following notation
\be
\begin{split}
& a_1 = z-1,\;\;\; a_2 = (1+x)/x,\;\;\; a_3 = (1+x(1-z))/x,\;\;\;\
\\
& 
a_4 = \frac{1 + y + x y + x y^2}{2 (1 + x)y},\;\;\;
a_5 = \frac{1 - y - x y + x y^2}{2 (1 + x)y},\;\;\;
a_6 = \frac{1+xy }{2 xy}.
\end{split}
\ee 

For  the three integrals that we show below, we separate real and imaginary parts and 
write 
\be
f^{NIJ}_{i} = {\rm Re} f^{NIJ}_{i} + i {\rm Im} f^{NIJ}_{i},
\ee

The explicit expressions read 
\be
\begin{split}
&{\rm Re} f^{N12}_{33} =  -\frac{1}{4}
+\ep \left[ \frac{1}{2} G\left(-\frac{1}{y},x\right)+\frac{1}{2} G(-1,x)-\frac{1}{2}
   G(0,y)+G(0,z)\right ]
\\
& +\ep^2 \Bigg [  G(-1,x) (G(0,y)-2 G(0,z))
+G(0,y)  \Bigg (-2 G\left(-\frac{z}{y},x\right)
+G\left(-\frac{1}{y},x\right)
\\
& 
+2 G(0,x)
-G(0,z)-G(1,z) \Bigg )
+G(0,z) \Big ( 2
   G\left(-\frac{z}{y},x\right)-2 G\left(-\frac{1}{y},x\right)
\end{split}
\ee
\be
\begin{split}
& 
+2 G\left(-\frac{1}{z},x\right)+G(z,y) \Big )
-2 G\left(-\frac{z}{y},0,x\right)-G\left(-1,-\frac{1}{y},x\right)
-G\left(-\frac{1}{y},-1,x\right)
\nn \\
& 
-G\left(-\frac{1}
   {y},-\frac{1}{y},x\right)
-2 G(0,x) G(0,z)+2 G\left(-\frac{1}{z},0,x\right)-G(-1,-1,x)+2 G(0,0,x)
\\
& 
+G(1,z)
   G(z,y)
-G(0,a_1,y)-G(z,0,y)+G(z,a_1,y)+2 G(0,0,y)-G(0,0,z) 
\\
& 
+G(0,1,z) 
+G(1,0,z)+G(1,1,z)-\frac{23 \pi
   ^2}{12}\Bigg ] + {\cal O}(\ep^3),
\\
& {\rm Im} f^{N12}_{33} = \frac{\pi \ep}{2} 
+\ep^2 \Bigg [ 2 \pi  G\left(-\frac{z}{y},x\right)-\pi  G\left(-\frac{1}{y},x\right)+2 \pi 
   G\left(-\frac{1}{z},x\right)-\pi  G(-1,x)
\\
& -2 \pi  G(0,x)+2 \pi  G(z,y)-3 \pi  G(0,y)+2 \pi  G(0,z)+2 \pi 
   G(1,z) \Bigg ] + {\cal O}(\ep^3).
\end{split}
\ee

\be
\begin{split}
& {\rm Re} f^{N13}_{33}  = 
\frac{1}{4}
-\frac{\ep}{2} \Big [ 
G\left(-\frac{1}{x},y\right)
+G(-1,x)-2 G(0,x)-G(a_1,y)-G(0,z)-G(1,z) \Big ]
\\
& -\ep^2 \Big [ G(-1,x) G(a_1,y)+G\left(-\frac{1}{x},y\right) (-G(-1,x)+2 G(0,x)+G(0,z)+G(1,z) )
 \\
& +G(1,z)
   \left(-2 G\left(-a_3,y\right)
+2 G(a_1,y)-G(z,y)\right)-2 G(0,x) G\left(-a_3,y\right)
\\
& 
+G\left(-\frac{1}{x},a_1,y\right)
+G\left(a_1,-\frac{1}{x},y\right)-2
   G\left(-a_3,a_1,y\right)
-G\left(-\frac{1}{x},-\frac{1}{x},y\right)
\\
& +G(-1,x) G(1,z)+G(-1,x) 
   G(0,z)-2 G(0,x) G\left(-\frac{1}{x},z\right)
-2 G(0,x) G\left(a_2,z\right)
\\
& 
-2 G\left(-\frac{1}{x},0,z\right)
 -2 G\left(a_2,1,z\right)
-G(-1,-1,x)+2 G(0,-1,x)
-G(0,z) G(z,y)
\\
& 
-G(a_1,0,y)+2 G(a_1,a_1,y)
+G(z,0,y)
-G(z,a_1,y)
+G(0,0,z)-G(0,1,z)
\\
& 
-G(1,0,z)+G(1,1,z)+3 \pi ^2 \Big ] +{\cal O}(\ep^3), 
\\
& {\rm Im} f^{N13}_{33}  = \frac{3\pi \ep }{2}
-\ep^2 \Big [ 
3 \pi G\left(-\frac{1}{x},y\right) 
-2 \pi  G\left(-a_3,y\right)
-\pi G(0,z)
-2  \pi  G\left(-\frac{1}{x},z\right)
\\
& 
-2  \pi  G\left(a_2,z\right)
+\pi  G(-1,x)
-2  \pi G(0,x)
+\pi  G(a_1,y)
-2  \pi  G(z,y)
-\pi  G(1,z)
\Big ] +{\cal O}(\ep^3). 
\end{split} 
\ee

\be
\begin{split}
& {\rm Re} f^{N34}_{37} = 
\ep^2 \Bigg [  
-G\left(a_6,a_4,z\right)+\frac{1}{2} G(0,a_4,z)+\frac{1}{2}
   G(1,a_4,z)
+\frac{1}{2} G\left(0,-a_5,z\right)
\\
& 
+\frac{1}{2} G\left(1,-a_5,z\right)
-G\left(-\frac{1-y}{2 y},-a_5,z\right)+G(0,x)
   \Big (G\left(-\frac{1-y}{2 y},z\right)
-G\left(a_6,z\right) \Big )
\end{split}
\ee
\be
\begin{split}
& +G(-1,y) \left(-G\left(\frac{x y+1}{2 x
   y},z\right)+G\left(\frac{1}{y},x\right)-G(0,x)
+G\left(-\frac{1-y}{2 y},z\right)\right)
\nn \\
& +G\left(\frac{1}{y},x\right)
   \left(G\left(a_6,z\right)
-G\left(-\frac{1-y}{2y},z\right)\right)
+G\left(0,\frac{1}{y},x\right)
+G\left(\frac{1}{y},0,x\right)
\\
& -G\left(\frac{1}{y},\frac{1}{y},x\right)-G
   (0,0,x)-G(-1,-1,y)-\frac{\pi ^2}{2} \Bigg ],
\\ 
&
{\rm Im} f^{N34}_{37} = 
 \ep^2 \Bigg [ -\pi  G\left(a_6,z\right)-\pi  G\left(-\frac{1}{y},x \right)+\pi  G(0,x)-\pi 
   G\left(-\frac{1-y}{2 y},z\right)
\\
& +2 \pi  G(0,y)-\pi  G(1,y)+\pi  G(0,z)+\pi  G(1,z)+2 \pi  \ln 2 \Bigg )
\end{split}
\ee

As a final comment,  we describe some checks  of our results. First, as we already mentioned, 
the boundary conditions for the integrals were obtained using two different methods. 
Second, we have checked that 
all computed integrals satisfy the differential equations. Third, many 
of the integrals that appear in this computation are, in fact, the planar ones. We have recalculated 
those integrals using the setup that is used for non-planar integrals, including different  
parametrization for the $N_{34}$ family, and found full agreement with our previous results. 
We computed some of the integrals {\it numerically} using the new version 
of the program FIESTA \cite{Smirnov:2013eza}, 
that is capable of calculating  certain Feynman integrals in the physical region. 
For all integrals  that were computed by FIESTA with sufficient accuracy, agreement 
with analytic results was found.  Finally, we compared  some integrals reported in this paper 
with the results of the recent calculation of two-loop four-point 
non-planar integrals in the equal mass case, reported 
recently in Ref.~\cite{Gehrmann:2014bfa}. For all the integrals compared, complete agreement was found.

\section{Conclusions} 
\label{sec:concl}

In this paper we reported on the computation of all two-loop non-planar master  integrals 
that are required to describe production of two {\it  off-shell} vector bosons in hadron collisions. 
These integrals were calculated using the differential equations method of 
Ref.~\cite{jhenn}. To solve the differential equations, we require boundary conditions. We computed 
the relevant boundary conditions   in the physical region and used them to 
construct analytic results for non-planar integrals 
in  terms of Goncharov polylogarithms. 

The results for the master integrals in terms of Goncharov polylogarithms,
as well as the matrices $\tilde{A}$ appearing in the differential equations, 
are included in the arXiv submission. We note that  representation of 
master integrals in terms of Goncharov polylogarithms may not be the most 
compact one but it 
has the advantage that these functions are by now
standard and
dedicated numerical implementations exist
\cite{Bauer:2000cp,Vollinga:2004sn}. Also, this representation manifestly
separates
real and imaginary parts.
We did not  try to simplify these results,
although such simplifications should be possible.
Probably the most compact and flexible form can be achieved in terms
of Chen iterated integrals \cite{Chen:1977oja}, at the cost of giving up the feature
of a linear parametrization. In this spirit, in the recent paper
\cite{Caron-Huot:2014lda} dealing with similar multi-scale integrals, it
was shown how a  one-dimensional integral representations can be
obtained that gives fast and reliable numerical results.
Another possibility is to rewrite the results in terms of a minimal function
basis (but allowing for more complicated arguments of those functions),
which up to weight four consists of classical polylogarithms and one other
function, which may be chosen to be ${\rm Li}_{2,2}$
\cite{Goncharov:2010jf,Duhr:2011zq}.

Finally, we note that the results presented in this paper provide the last missing ingredient -- the non-planar 
master integrals -- for the computation of  two-loop amplitudes that describe annihilation 
of two massless partons into two off-shell  gauge bosons.  Once these amplitudes 
become available,  theoretical predictions for the production of electroweak gauge bosons at the LHC 
will be substantially improved. 

\section*{Acknowledgments} 

We would like thank L.~Tancredi for providing numerical cross-checks for some of the results 
reported in this paper. 
J.M.H. is supported in part by
the DOE grant DE-SC0009988 and by the Marvin L. Goldberger fund.
The work of F.C. and K.M.   is partially supported by US NSF under grant PHY-1214000.
K.M. is also   supported by Karlsruhe Institute of Technology through its distinguished 
researcher fellowship program.  
The work of V.S. was supported by the Alexander von Humboldt Foundation (Humboldt Forschungspreis).
We are grateful to the Institute for Theoretical Particle Physics (TTP) at Karlsruhe Institute of Technology 
where some of the results were obtained. 
We are indebted  to A.~Smirnov for the possibility to use his
{\tt c++} version of {\tt FIRE}.

\vskip 0.5cm
{\bf Note added} We are grateful to Andreas von Manteuffel for pointing out a few misprints in the
original version of the paper.

\end{document}